\documentclass{article}
\usepackage[utf8]{inputenc}
\usepackage[top=1in, left=1in, right=1in, bottom=1in]{geometry}
\usepackage{amsmath}
\usepackage{amssymb}
\usepackage{enumitem}
\usepackage{graphicx}
\usepackage{float}
\usepackage{cite}
\usepackage{subcaption}
\usepackage{xcolor}
\usepackage[makeroom]{cancel}
\usepackage{svg}
\usepackage[title]{appendix}
\usepackage{todonotes}
\usepackage{booktabs}

\title{Time-discretization of a plasma-neutral MHD model with a semi-implicit leapfrog algorithm}
\author{Sina Taheri, Jacob R King, Uri Shumlak}
\vspace{-2ex}

\begin{document}

\maketitle

\begin{abstract}
    The semi-implicit leapfrog time-discretization is a workhorse algorithm for initial-value MHD codes to bridge between vastly separated time scales. Inclusion of atomic interactions with neutrals breaks the functional structure of the MHD equations that exploited by the leapfrog. We address how to best integrate atomic physics into the semi-implicit leapfrog. Following the Crank-Nicolson method, one approach is to time-center the atomic interactions in the linear solver and use a Newton method to include the nonlinear contributions. Alternatively, another family of methods are based on operator-splitting the terms associated with the atomic interactions using a Strang-splitting technique. These methods naturally break equations into constituent ODE and PDE parts and preserve the structure exploited by the semi-implicit leapfrog. We study the accuracy and efficiency of these methods through a battery of 0D and 1D cases and show that a second-order-in-time Douglas-Rachford inspired coupling between the ODE and PDE advances is effective in reducing the time-discretization error to be comparable to that of Crank-Nicolson with Newton iteration of the nonlinear terms. Splitting ODE and PDE parts results in independent matrix solves for each field which reduces the computational cost considerably and provides parallelization over species relative to Crank-Nicolson. 
\end{abstract}

\section{Introduction} 
A key challenge to simulating magnetized-plasma dynamics, as governed by the magnetohydrodynamic (MHD) equations~\cite{Schnack2009}, is incorporating the disparate time scales. Fast dynamics are mediated by Alfv\'en waves while changes to the magnetic topology occurs on a slower time scale determined by the low plasma resistivity. Atomic interactions are important but often neglected in MHD modeling. Fluid equations for atomic interactions introduce additional stiff coupling through effects such as ionization, recombination, and charge exchange.

Without atomic interactions, discretization of the MHD equations using a mixed implicit/semi-implicit leapfrog method~\cite{Schnack1987,Sovinec2004,Sovinec2010} is often employed to bridge the fast-wave and slow-diffusive time scales. By exploiting the functional structure of the MHD equations, the advances for the plasma density, velocity, temperature and magnetic field can be staggered or solved sequentially which results in a smaller number of degrees of freedom during each separated field advance relative to the full system of equations. As the computationally cost to solve an implicit system scales greater than linearly with the degrees of freedom this leads to an efficient time discretization. In this work, we investigate how to best integrate atomic physics into the semi-implicit leapfrog time discretization within the NIMROD code~\cite{Sovinec2004,Sovinec2010}. The inclusion of atomic interactions breaks the functional structure of the MHD equations that is exploited by this discretization. We specifically address the most basic limiting case: a hydrogenic species with both neutral and ionized components~\cite{Meier2012}. These model equations are relevant to a wide range of applications such as low-temperature plasmas~\cite{diomede2008fluid}, astrophysical plasmas~\cite{khomenko2014fluid}, plasma formation~\cite{dunlea2020model}, and fusion-plasma edge fueling~\cite{casali2020neutral}.

We consider two competing approaches for this implementation: coupling the neutral species to the plasma through a time-centered Crank-Nicolson approach within the structure of the leapfrog advance versus introducing additional operator-splitting for the atomic physics interactions. Regarding the Crank-Nicolson approach, the coupled neutral and plasma density, velocity and temperature are each solved simultaneously but coupling between some fields (e.g. centering the temperature in the density advance) is neglected as consistent with the leapfrog staggering. One challenge to this approach is including full time centering for the nonlinear terms that arise through the atomic interactions. By employing an approximate Jacobian-free Newton-Krylov (JFNK) method~\cite{Kelley2003,Knoll2004} we are able to solve with full nonlinearities for each field; however, the associated iteration increases the computational cost. We consider both linear-implicit and full JFNK (implicitly balanced) variations of this method. Using asymptotic analysis of considerably simpler radiation-diffusion system, Ref.~\cite{Knoll2006} shows that a linear-implicit method can produce erroneous answers at large time-step size relative to the dynamical-system time. However, an implicitly balanced method preserves the asymptotic-equilibrium solution exactly.

Alternatively, operator-splitting the terms associated with the atomic interactions using a Strang-splitting technique~\cite{Strang1968} naturally breaks equations into constituent ODE (no spatial gradients) and PDE parts (which capture all the spatial gradients and are unchanged from the MHD equations without atomic interactions). Since all of the nonlinear atomic interactions are naturally handled by an ODE solver, no nonlinear iteration is required. 
Furthermore, we consider two different operator-split formulations: Strang-splitting with half-step field-specific ODE solves surrounding each field update in the leapfrog advance and Strang-splitting with half-step full-equation-system ODE solves interleaved into the leapfrog advance combined with a Douglas-Rachford inspired coupling~\cite{Douglas1956} between the ODE and PDE advances. The latter formulation both eliminates the error associated with the leapfrog structure neglecting the functional dependencies introduced by the atomic physics terms and reduces the ODE/PDE splitting error.

These discretization approaches are unique relative to other codes that solve coupled plasma-neutral equations --- none of which employ leapfrog splitting of the MHD equations. For 2D fusion-plasma edge applications, SOLPS and UEDGE are two prominent codes. SOLPS combines a 2D multi-fluid plasma transport code, with a 3D kinetic Monte Carlo neutral transport code~\cite{Wiesen2015}. The fluid code is based on a finite volume discretization of the conservation equations and is fully implicit in time~\cite{braams1987multi}. The model consists of continuity equations and equations for parallel momentum balance for all charge states of the bulk species, in addition to an electron temperature equation, an ion+neutral temperature equation, and an electric potential equation. Parallel transport is classical according to Braginskii and Balescu, though flux-limited; perpendicular transport is anomalous~\cite{braams1996radiative}. The UEDGE code, on the other hand, is a time dependent 2D finite difference fluid code~\cite{rognlien1992fully,rognlien2017comparison}. It solves the classical Braginskii transport equations for plasma density, parallel momentum and thermal ion and electron energy transport along the magnetic field, and assumes anomalous diffusive transport across the field. The code uses a  reduced Navier-Stokes model for atomic neutrals with a single momentum equation for transport along field lines~\cite{rensink1999edge}. 

There are other fluid based codes, e.g., HiFi~\cite{meier2010spectral}, PSI-Tet~\cite{hansen2015simulation}, and WARPXM~\cite{shumlak2011advanced}, that are using the same plasma-neutral model as described here with different time discretization schemes. HiFi uses a spectral element method with fully implicit time discretization to solve single temperature plasma equations coupled with a neutral fluid. The full Jacobian of the system of equations is required for Crank-Nicolson time-centering method in HiFi. PSI-Tet is a finite element two-fluid extended MHD code that uses fully implicit Crank-Nicolson algorithm for time advance. PSI-Tet uses a Newton-Raphson method solve the system of nonlinear equations with a Krylov-space iterative method. Finally, WARPXM is a discontinuous-Galerkin finite-element code with an explicit strong-stability-preserving (SSP) time-advance method. Jacobian calculation is not required in WARPXM; however, preserving stability requires small time-step sizes.  

In Sec.~\ref{sec:model} we describe the model equations and associated approximations in more detail and then describe the competing time-discretization schemes. The discussion next turns to diagnosing the accuracy and computationally efficiency of each time-discretization by quantifying the error from different sources (e.g. from using a leapfrog method with atomic interactions, neglect of the nonlinear contributions and operator splitting) and number of solves associated with each method. In Sec.~\ref{sec:test0-D} we employ a number of zero-dimensional tests with varied plasma-neutral parameters. Since these tests neglect any error associated with the ODE-PDE splitting we move on to one-dimensional tests in Sec.~\ref{sec:test1-D} by considering cases that roughly represent tokamak-plasma edge fueling and z-pinch startup from a plasma-gun injector.

\section{Model} \label{sec:model}
The NIMROD MHD code~\cite{Sovinec2004} uses 2D nodal spectral elements combined with a 1D Fourier representation to solve 3D problems. The MHD equations are supplemented with the dynamic neutral model presented in Ref.~\cite{Meier2012} which incorporates electron impact ionization, radiative recombination and resonant charge-exchange. In this model plasma velocity is described by the center-of-mass velocity for ions and electrons and a neutral velocity is introduced for neutral species. The coupled system of equations is then

\subsubsection*{number densities}
\begin{equation} \label{eq:plasma_nd}
    \frac{\partial n}{\partial t} + \nabla \cdot (n \textbf{V}) = \Gamma^{ion} - \Gamma^{rec}
\end{equation}
\begin{equation} \label{eq:neutral_nd}
    \frac{\partial n_n}{\partial t} + \nabla \cdot (n_n \textbf{V}_n) = \Gamma^{rec} - \Gamma^{ion}
\end{equation}

\subsubsection*{velocities}
\begin{equation} \label{eq:plasma_com_vel}
    m n \left( \frac{\partial}{\partial t} +  \textbf{V} \cdot \nabla \right)  \textbf{V} = \textbf{J} \times \textbf{B} - \nabla p - \nabla \cdot \underbar{$\Pi$} + \textbf{R}_{in}^{cx} - \textbf{R}_{ni}^{cx} + m (\Gamma^{ion} + \Gamma^{cx}) (\textbf{V}_n - \textbf{V})
\end{equation}
\begin{equation} \label{eq:neutral_vel}
    m_n n_n \left( \frac{\partial}{\partial t} +  \textbf{V}_n \cdot \nabla \right)  \textbf{V}_n = - \nabla p_n - \nabla \cdot \underbar{$\Pi$}_n + \textbf{R}_{ni}^{cx} - \textbf{R}_{in}^{cx} + m (\Gamma^{rec} + \Gamma^{cx}) (\textbf{V} - \textbf{V}_n)
\end{equation}

\subsubsection*{temperatures}
\begin{equation} \label{eq:plasma_temp}
\begin{split}
    \frac{n}{\Gamma - 1} \left( \frac{\partial}{\partial t} + \textbf{V} \cdot \nabla \right) T =
    &- n T \nabla \cdot \textbf{V} - \nabla \cdot \textbf{q} + Q + Q^{ion} - Q^{rec} + Q_{in}^{cx} - Q_{ni}^{cx} - \Gamma^{ion} \phi_{ion} \\
    &- \frac{1}{\Gamma - 1} T (\Gamma^{ion} - \Gamma^{rec}) + \frac{m}{2} (\Gamma^{ion} + \Gamma^{cx}) (\textbf{V} - \textbf{V}_n)^2 - \textbf{R}_{in}^{cx} \cdot (\textbf{V} - \textbf{V}_n)
\end{split}
\end{equation}
\begin{equation} \label{eq:neutral_temp}
\begin{split}
    \frac{n_n}{\Gamma - 1} \left( \frac{\partial}{\partial t} + \textbf{V}_n \cdot \nabla \right) T_n = &- n_n T_n \nabla \cdot \textbf{V}_n - \nabla \cdot \textbf{q}_n + Q_n + Q^{rec} - Q^{ion} + Q_{ni}^{cx} - Q_{in}^{cx} \\
    &- \frac{1}{\Gamma - 1} T_n (\Gamma^{rec} - \Gamma^{ion}) + \frac{m}{2} (\Gamma^{rec} + \Gamma^{cx}) (\textbf{V} - \textbf{V}_n)^2 + \textbf{R}_{ni}^{cx} \cdot (\textbf{V} - \textbf{V}_n)
\end{split}
\end{equation}
where $\textbf{V}$ is the plasma center-of-mass velocity, $\textbf{V}_n$ is the neutrals velocity, $\textbf{q}$ and $\textbf{q}_n$ are conductive-heat flux for plasma and neutrals, $Q$ and $Q_n$ are the scattering collisional heat exchanges, $Q^{ion}$, $Q^{rec}$, and $Q_*^{cx}$ are the ionization, recombination, and charge-exchange heat exchanges, respectively, $\textbf{R}_*^{cx}$ is the frictional momentum transfer, $\Gamma^{ion}$, $\Gamma^{rec}$, and $\Gamma^{cx}$ are number-density source rates due to ionization, recombination, and charge-exchange reactions, and $\phi_{ion}$ is the effective ionization energy. These closures, detailed in Ref.~\cite{Meier2012}, are summarized as
\begin{equation} \label{eq:ion_src}
    \Gamma^{ion} = \langle \sigma_{ion} v \rangle n n_n
\end{equation}
\begin{equation} \label{eq:rec_src}
    \Gamma^{rec} = \langle \sigma_{rec} v \rangle n^2
\end{equation}
\begin{equation} \label{eq:cx_src}
    \Gamma^{cx} = \sigma_{cx} n n_n \sqrt{\frac{4}{\pi} v_{Ti}^2 + \frac{4}{\pi} v_{Tn}^2 + v_{in}^2}
\end{equation}
\begin{equation} \label{eq:cx_momontum_in}
    \textbf{R}_{in}^{cx} \approx - m \sigma_{cx} n n_n (\textbf{V} - \textbf{V}_n) v_{Tn}^2 \left[ 4 \left( \frac{4}{\pi} v_{Ti}^2 + v_{in}^2 \right) + \frac{9 \pi}{4} v_{Tn}^2 \right]^{-1/2}
\end{equation}
\begin{equation} \label{eq:cx_momontum_ni}
    \textbf{R}_{ni}^{cx} \approx + m \sigma_{cx} n n_n (\textbf{V} - \textbf{V}_n) v_{Ti}^2 \left[ 4 \left( \frac{4}{\pi} v_{Tn}^2 + v_{in}^2 \right) + \frac{9 \pi}{4} v_{Ti}^2 \right]^{-1/2}
\end{equation}
\begin{equation} \label{eq:ion_heatx}
    Q^{ion} = \frac{3}{2} T_n \Gamma^{ion}
\end{equation}
\begin{equation} \label{eq:rec_heatx}
    Q^{rec} = \frac{3}{2} T_i \Gamma^{rec}
\end{equation}
\begin{equation} \label{eq:cx_heatx_in}
    Q_{in}^{cx} \approx \frac{3}{4} m \sigma_{cx} n n_n v_{Tn}^2 \sqrt{\frac{4}{\pi} v_{Ti}^2 + \frac{64}{9\pi} v_{Tn}^2 + v_{in}^2}
\end{equation}
\begin{equation} \label{eq:cx_heatx_ni}
    Q_{ni}^{cx} \approx \frac{3}{4} m \sigma_{cx} n n_n v_{Ti}^2 \sqrt{\frac{4}{\pi} v_{Tn}^2 + \frac{64}{9\pi} v_{Ti}^2 + v_{in}^2}
\end{equation}
where $v_{T_*}^2 = 2 T_* / m$ is the thermal velocity and $v_{in}^2 = \| \textbf{V} - \textbf{V}_n \|^2$ is the square of the relative velocity between plasma and neutrals. The ionization and recombination cross-section rates from the DEGAS2 code are used~\cite{Stotler1994}.

Some approximations are made regarding the atomic physics. The plasma-neutral model includes only singly ionized hydrogenic plasmas where the excited states for neutral atoms are not tracked; however, this is accounted for through the DEGAS2 combined cross-section rates. Also, we assume an optically thin plasma-neutral system with prompt loss of energy due to radiative atomic processes. We neglect the neutral-charged particle collisions compared to Coulomb collisions as is appropriate for plasmas that are even a few percent ionized \cite{Goldston1995}. Thus, the scattering collisional transfer of momentum between plasma and neutrals is dropped. Consistently, the scattering collisional heat generation due to neutral-charged particle collisions is neglected in plasma and neutral energy equations.  

Next we discuss the closures of the higher order moments. The total plasma stress tensor is $\underbar{$\Pi$} = \underbar{$\Pi$}_i + \underbar{$\Pi$}_e$ where electron stress tensor is much smaller than ion stress tensor, so the approximation $\underbar{$\Pi$} \approx \underbar{$\Pi$}_i$ is appropriate based on the electron-to-ion-mass ratio~\cite{Meier2012}. We assume isotropic stress tensors and heat fluxes for both plasma and neutral species given by
\begin{equation} \label{eq:stress_tensor}
    \underbar{$\Pi$}_* = -m n \chi_* \left[ \nabla \textbf{V}_* + \nabla \textbf{V}_*^T - \frac{2}{3} \mathbb{I} \nabla \cdot \textbf{V}_* \right]
\end{equation}
\begin{equation} \label{heat_flux}
    \textbf{q}_* = - \kappa_* \nabla T_*
\end{equation}
where $\chi_*$ is the viscous diffusivity and $\kappa_*$ is the thermal conductivity. While NIMROD is fully capable of including highly anisotropic and finite gyro radius effects in the closures, these are not a focus of this study. We do not anticipate numerical complications from the interactions with the neutral particles associated with their usage.

Finally, the magnetic field, $\textbf{B}$, and the current density, $\textbf{J}$, are related by the low-frequency Ampere's law, $\nabla \times \textbf{B} = \mu_0 \textbf{J}$. Combining the Faraday's law with the Ohm's law, we write the magnetic induction equation as 
\begin{equation} \label{eq:induction}
    \frac{\partial \textbf{B}}{\partial t} = - \nabla \times \left[ \eta \textbf{J} - \textbf{V} \times \textbf{B} \right].
\end{equation}
The NIMROD code is capable of using a generalized Ohm's law accounting for Hall effect, neoclassical effects, and electron pressure term. In this study, we do not exercise the full generalized Ohm's law. As appropriate for study of low-frequency dynamics, we neglect the displacement current and assume quasi-neutrality. Consistent with the discussion above, we drop the electron-neutral scattering term.

\section{Time Discretization} \label{sec:time-discretization}

Consider the functional dependencies for single-fluid MHD without atomic interactions: $n(\textbf{V})$, $\textbf{V}(n,\textbf{B},T)$, $T(n,\textbf{V})$, and $\textbf{B}(\textbf{V})$ from Eqns.~\eqref{eq:plasma_nd}, \eqref{eq:plasma_com_vel}, \eqref{eq:plasma_temp}, \eqref{eq:induction}. This data dependence allows for time-splitting by field and sequential field solves which reduces the complexity and computational cost of implicit algorithms. In the semi-implicit leapfrog algorithm the velocity field is staggered with respect to the other fields which are solved sequentially~\cite{Sovinec2004}. In order to stabilize wave propagation associated with the leapfrog algorithm, it is essential to include a semi-implicit operator~\cite{Schnack1987} which is added to the velocity advance equation in the NIMROD code. Using the ideal-MHD force operator as the semi-implicit operator limits spectral-pollution time-discretization error~\cite{Lutjens1996}. In this work, we consider extension to the NIMROD mixed implicit/semi-implicit leapfrog method presented in Ref.~\cite{Sovinec2010} and do not address the predictor-corrector algorithms~\cite{Sovinec2004}. Analogously, a time-staggered algorithm for a neutral species also requires a semi-implicit operator to stabilize the sound-wave propagation. Details of the operators used in this study are given in Appendix~\ref{sec:appendixA}.

In the following subsections, we discuss competing time-discretization algorithms. Section~\ref{subsec:CN-Linear} details an algorithm (CN-linear) with Crank-Nicolson time centering of the linearized contributions from the atomic physics terms. In Sec.~\ref{subsec:CN-nonlinear}, we describe Newton iteration that adds full nonlinearity to the Crank-Nicolson time centering as required for an accurate and stable algorithm (CN-JFNK). As a different approach, Secs.~\ref{subsec:SS-multispecies} and~\ref{subsec:DR-multispecies} gives details of the operator-splitting algorithms, SS-field and SS-leapfrog+DR, respectively. Comparison of these different algorithms is a key purpose of this paper.

\subsection{Crank-Nicolson time-centering within semi-implicit leapfrog algorithm} \label{subsec:CN-Linear}
Following Refs.~\cite{Sovinec2004} and~\cite{Sovinec2010} , the discrete form of an implicit/semi-implicit numerical time-advance for the number density equation of the plasma-neutral model is

\begin{equation} \label{eq:nd_discrete}
\begin{split}
	\frac{\Delta n}{\Delta t} &+ \theta \nabla \cdot [\textbf{V}^{j+1} \Delta n] - \theta \Gamma^{ion}_1 (\Delta n, \Delta n_n, n^{j+\frac{1}{2}}, n_n^{j+\frac{1}{2}}, T^{j+\frac{1}{2}}) + \theta \Gamma^{rec}_1 (\Delta n, n^{j+\frac{1}{2}}, T^{j+\frac{1}{2}}) \\
	&= - \nabla \cdot [\textbf{V}^{j+1} n^{j+\frac{1}{2}}] + \Gamma^{ion}_0 (n^{j+\frac{1}{2}}, n_n^{j+\frac{1}{2}}, T^{j+\frac{1}{2}}) - \Gamma^{rec}_0 (n^{j+\frac{1}{2}}, T^{j+\frac{1}{2}})
\end{split}
\end{equation}

\begin{equation} \label{nn_discrete}
\begin{split}
	\frac{\Delta n_n}{\Delta t} &+ \theta \nabla \cdot [\textbf{V}_n^{j+1} \Delta n_n] + \theta \Gamma^{ion}_1 (\Delta n, \Delta n_n, n^{j+\frac{1}{2}}, n_n^{j+\frac{1}{2}}, T^{j+\frac{1}{2}}) - \theta \Gamma^{rec}_1 (\Delta n, n^{j+\frac{1}{2}}, T^{j+\frac{1}{2}}) \\
	&= - \nabla \cdot [\textbf{V}_n^{j+1} n_n^{j+\frac{1}{2}}] - \Gamma^{ion}_0 (n^{j+\frac{1}{2}}, n_n^{j+\frac{1}{2}}, T^{j+\frac{1}{2}}) + \Gamma^{rec}_0 (n^{j+\frac{1}{2}}, T^{j+\frac{1}{2}})
\end{split}
\end{equation}
where superscript shows the discrete step label, $\Delta n$ is the change in number density, $n^{j+3/2} - n^{j+1/2}$, and $\theta$ is the centering parameter. $\textbf{V}$ is centered in the density advance through the leapfrog algorithm. In this work Crank-Nicolson centering of $\theta=1/2$ is used for the hyperbolic and atomic reaction terms. Reference~\cite{Sovinec2010} shows that time centering of $\theta=1/2$ for advection terms is required to maintain numerical stability for the MHD advance with the mixed implicit/semi-implicit leapfrog. 

Equations \eqref{eq:nd_discrete} and \eqref{nn_discrete} are coupled through atomic reactions between plasma and neutrals. To clearly present the discrete form of atomic physics we expand the ionization and recombination source terms around a time-step as $\Gamma^* (n + \theta \Delta n)= \Gamma^*_0 (n) + \theta \Gamma^*_1 (n, \Delta n, \theta) + \Gamma^*_{nl} (n, \Delta n, \theta)$ where $\Gamma^*_0$ is the full source term at the beginning of time-step, $\Gamma^*_1$ is the linear contribution of atomic physics due to change in the field, and $\Gamma^*_{nl}$ represents the nonlinear contribution of the field change in atomic physics source terms. For example, the expansion of the ionization source 
term in plasma number density equation can be presented as
\begin{equation} \label{eq:ionization_expand}
    \Gamma^{ion} (n + \theta \Delta n, n_n + \theta \Delta n_n) = \langle \sigma_{ion} v \rangle n n_n + \theta \langle \sigma_{ion} v \rangle \Delta n \ n_n + \theta \langle \sigma_{ion} v \rangle n \ \Delta n_n + \Gamma^{ion}_{nl}
\end{equation}
where $\Gamma^{ion}_0 = \langle \sigma_{ion} v \rangle n n_n$, $\Gamma^{ion}_1 = \langle \sigma_{ion} v \rangle \Delta n \ n_n + \langle \sigma_{ion} v \rangle n \ \Delta n_n$, and $\Gamma^{ion}_{nl} = \Gamma^{ion} - \Gamma^{ion}_0 - \theta \Gamma^{ion}_1$. The $\Gamma^*_1$ term does not include a linear contribution from the cross-section, $\langle \sigma v \rangle$, which may be density dependent. This choice is made as the functional form of the cross section may not be known a priori. These contributions are included in $\Gamma^*_{nl}$ which will be addressed in Sec.~\ref{subsec:CN-nonlinear}.

\begin{figure}
    \centering
    \includesvg[scale=0.6]{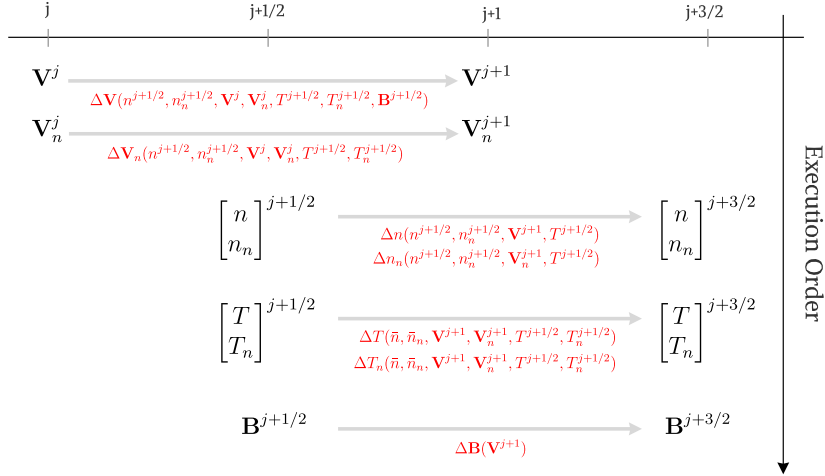}
    \caption{Schematic of Crank-Nicolson time advance. By exploiting the functional structure of MHD equations, velocity, number density, temperature, and magnetic field advances can be staggered and solved sequentially using the semi-implicit leapfrog algorithm. Plasma and neutral velocities are solved separately where number densities and temperatures for both species are solved simultaneously.}
    \label{fig:CN-scheme}
\end{figure}

Turning to the full system of equations, available in Appendix~\ref{sec:appendixA}, Fig.~\ref{fig:CN-scheme} shows the time staggering and data dependency schematically. Velocity is staggered in time with respect to number density, temperature, and magnetic field which are solved sequentially. The ordering of the sequential solves is based on the MHD data dependencies. However, including neutral-plasma atomic interactions breaks this data dependency between number density and temperature equations when the cross-section depends on temperature as is common. Explicit centering of temperature in the number density equation amplifies the time-discretization error associated with temperature-dependent cross-sections. We refer to this error as \textit{T-in-n} error. 

While the linearized velocity fields for plasma and neutral species are solved independently here, we add full CN time-centering to them through nonlinear terms in Newton iteration as detailed in Sec.~\ref{subsec:CN-nonlinear}. Square bracket notation in Fig.~\ref{fig:CN-scheme} implies that the species are dependent on each other for the specified field and are solved simultaneously. Although we could solve the number density and temperature equations simultaneously to eliminate the \textit{T-in-n} error, we choose not to due to implementation complexity, computational cost and potential issues with system-size scaling with the number of atomic species.

\subsection{Newton iteration for nonlinear terms in Crank-Nicolson method} \label{subsec:CN-nonlinear}
We now turn to how a Newton method is applied for the nonlinearities (and a small subset of linearized contributions as mentioned in Sec.~\ref{subsec:CN-Linear}). The nonlinear dynamics of plasma-neutral interactions at low temperature plasmas ($\sim 1-5$ eV) mostly depend on the atomic reaction cross-section rates which are calculated by DEGAS2 code and computed via bi-cubic spline interpolation from a table. The Newton method is a straight forward algorithm to include nonlinear terms in the time advance.  

For completeness we briefly summarize Newton's method presented in Ref.~\cite{Kelley2003}, while also highlighting specific implementation details. Consider a nonlinear system of equations given as $dx/dt = g(x)$, where $g(x)$ is a general nonlinear function of $x$. We discretize the equation using the Crank-Nicolson time-centering scheme. Expanding the nonlinear function with Taylor series, we get
\begin{equation}\label{eq:nl_newton}
    \Delta x - \Delta t \left[ g(x^n) + \frac{1}{2} \Delta x \ g'(x^n) + g_{nl}(x^n, \Delta x) \right] = 0,
\end{equation}
where prime notation presents the derivative with respect to variable $x$. $g_{nl}$ incorporates all the high order terms in Taylor expansion and can be simply calculated using the Taylor expansion definition,
\begin{equation} \label{eq:g_nl}
    g_{nl}(x^n, \Delta x) = g(x^n + \frac{1}{2} \Delta x) - g(x^n) - \frac{1}{2} \Delta x \ g'(x^n).
\end{equation}
Considering that $g(x)$ can be a system of equations, we expand the definition of $g_{nl}$ to include the inter-species interaction terms from $g'(x^n)$, cross-section derivatives, and the nonlinear part of the Taylor expansion. The inter-species velocity advance equations (Eqs.~\eqref{eq:app_vcom}-\eqref{eq:app_ntemp} in Appendix~\ref{sec:appendixA}) are CN-time-centered through nonlinear Newton iteration eliminating the limitation discussed in linear Crank-Nicolson algorithm in Sec.~\ref{subsec:CN-Linear}. Independent linear solve of the velocity fields avoids the need for a 6-component vector solve which reduces the complexity and cost of the numerical solver. 

A Newton-Krylov method is used to solve the linear part of the system of equations. After simplifying and rearranging Eq.~\eqref{eq:nl_newton}, the Newton sequence to find the root is given as
\begin{equation} \label{eq:newton_sequence}
    \left[ 1 - \frac{1}{2} \Delta t \ g'(x^n) - \Delta t \ \frac{\partial g_{nl} (x^n, \Delta x_k) }{\partial \Delta x_k} \right] \Delta x_{k+1} = \Delta t \left[ g(x^n) + g_{nl}(x^n, \Delta x_k) - \Delta x_k \frac{\partial g_{nl}(x^n, \Delta x_k)}{\partial \Delta x_k} \right],
\end{equation}
where the partial derivative with respect to $\Delta x$ is the Jacobian of the high order terms with respect to the change in the variable, $x^n$ represents the variables at time step $n$, and $\Delta x_k$ is the change in the variables at iteration $k$. Since $g_{nl}(x^n, \Delta x_k)$ has no analytic form with tabular atomic-physics cross sections, a simple finite-difference approximation is used to find its Jacobian as
\begin{equation} \label{g_nl_jacobian}
    \frac{\partial g_{nl}(x, \Delta x)}{\partial \Delta x} = \frac{g_{nl}(x, \Delta x) - g_{nl}(x, (1 - \epsilon) \Delta x)}{\epsilon \Delta x}.
\end{equation}

An error estimate is required to test convergence, and in most cases the norm of Eq.~\eqref{eq:nl_newton} can be used to show the decay rate in error~\cite{Kelley2003}. Accordingly, the convergence criterion considered for the nonlinear Newton iteration is 
\begin{equation} \label{eq:nl_tolerance}
    \frac{\| [1 - \frac{1}{2} \Delta t \ g'(x^n)] \Delta x^k - \Delta t \ g(x^n) - \Delta t \ g_{nl}(x^n, \Delta x^k) \|}{\| \Delta t \ g(x^n) \|} < tol_{nl},
\end{equation}
where $tol_{nl}$ is the nonlinear solver tolerance. 

At large time-step size, linearized operators can cause large overshoots relative to the solution during nonlinear iteration. Although the numerical schemes can recover the overshoot and find the correct steady-state solution, in fluid dynamics simulations these overshoots can cause physically impossible solutions such as negative pressure which often terminates the simulation. A ``dynamic" floor is implemented to keep number density and temperature positive during iteration. Another knob used to control the Newton nonlinear iterations is relaxing the solution, in which a relaxation factor limits how far the new guess for $\Delta x$ can be from the previous solution. 

\subsection{Strang-split time advance within the semi-implicit leapfrog algorithm} \label{subsec:SS-multispecies}
The semi-implicit leapfrog with Crank-Nicolson centering handles the MHD-fluid dynamics and atomic reactions simultaneously. An alternative approach is operator splitting the atomic physics from fluid dynamics. Additional operator splitting is in the spirit of leapfrog algorithm. In this method the spatially local atomic reactions (ODE system) is advanced half time-step. Then the MHD-fluid equations (PDE system) evolves a full time-step using the semi-implicit leapfrog algorithm. Finally, the ODE system is advanced for another half time-step to complete the Strang-split algorithm. General form of the iteration for this method is given by  
\begin{equation} \label{eq:strang-split}
    \begin{split}
        x^{(j+1, j+1/2)} &= \mathcal{S}_{\text{ode}} \bigg(\frac{\Delta t}{2}, x^{(j+1/2, j+1/2)} \bigg), \\
        x^{(j+1, j+3/2)} &= \mathcal{S}_{\text{pde}} \bigg(\Delta t, x^{(j+1, j+1/2)} \bigg), \\
        x^{(j+3/2, j+3/2)} &= \mathcal{S}_{\text{ode}} \bigg(\frac{\Delta t}{2}, x^{(j+1, j+3/2)} \bigg),
    \end{split}
\end{equation}
where $x$ is one of fields $n$, $\textbf{V}$, or $T$,  $\mathcal{S}_{\text{ode}}$ is the numerical solution for the ODE system composed of the atomic reactions between species, and $\mathcal{S}_{\text{pde}}$ represents the numerical solution for the MHD-fluid system. The first superscript represents the ODE time index, where the second one denotes the time indexing for the PDE contribution. 

Consider the plasma density equation, Eq.~\eqref{eq:plasma_nd}, as an example. Operator splitting divides the solution into two parts: one part is the solution to the ODE system given by
\begin{equation} \label{eq:ODE}
    \frac{\partial n}{\partial t} = \Gamma^{ion} - \Gamma^{rec},
\end{equation}
and the other part is the solution to the PDE system given by the nonlinear equation
\begin{equation} \label{eq:PDE}
    \frac{\partial n}{\partial t} + \nabla \cdot (n \textbf{V}) = 0. 
\end{equation}

In this algorithm the coupling between species is captured in the ODE system of equations. Within the scope of the semi-implicit leapfrog algorithm, the Strang operator splitting is applied to $n$, $\textbf{V}$ and $T$ separately. 
\begin{figure}
    \centering
    \includesvg[scale=0.6]{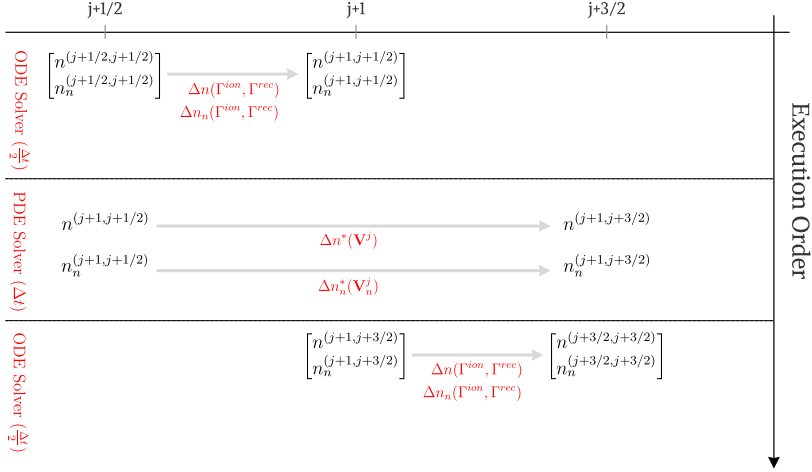}
    \caption{Schematic of SS-field time advance for number density equation. Local atomic physics are captured by the ODE system of equations where the PDE system represents the MHD-fluid equations for plasma and neutral species. Since plasma-neutral interactions are captured in local ODEs, the PDE equations can be solved independently.}
    \label{fig:SS-scheme}
\end{figure}

Figure~\ref{fig:SS-scheme} shows a schematic view of the SS-field time advance for plasma and neutral number density equations. Away from the equilibrium point the atomic reactions are local processes that happen on much faster time-scales compared to typical MHD-fluid waves which makes the ODE problem stiff. There are plenty of high quality methods developed for this type of problems. Here we use a stiff/non-stiff ODE solver, \texttt{lsode}, with Adams methods~\cite{lsode-llnl}. Since plasma and neutral species are only coupled through atomic reactions, operator splitting makes the fields in the MHD-fluid dynamics independent. As a result, the global sparse matrices for the implicit solve become smaller, reducing the numerical cost for the PDE part of the problem. 

Field separation and sequential solve is still in the SS-field method, thus it does not resolve the \textit{T-in-n} error. All the nonlinearities in the atomic physics are captured in the ODE solver, thus SS-field method eliminates any error related to nonlinear dynamics in the source terms. 

\subsection{Strang-split time advance within the leapfrog with Douglas-Rachford inspired coupling} \label{subsec:DR-multispecies}
Now we focus on an alternative splitting method which improves the SS-field method in two aspects. First, the Strang-split is interleaved into the leapfrog where all fields are integrated simultaneously by the ODE solver which eliminates the \textit{T-in-n} error. Second, incorporation of a correction source term inspired by Douglas-Rachford coupling~\cite{Douglas1956} that reduces the error associated with ODE/PDE splitting. We will discuss more details about different errors in Sec.~\ref{subsec:error_source}. Consider the full system of plasma-neutral equations as
\begin{equation} \label{eq:linearized_model}
    \frac{d\textbf{u}}{dt} = \mathcal{A}_1 (\textbf{u}) + \mathcal{A}_2 (\textbf{u}),
\end{equation}
where $\mathcal{A}_1$ is the generalized form of Eq.~\eqref{eq:ODE} for all the fields in which the ODE equations describe the nonlinear local atomic physics and $\mathcal{A}_2$ is the generalized form of Eq.~\eqref{eq:PDE} representing the linearized MHD-fluid system of equations. Adding Douglas-Rachford inspired coupling with a Strang-splitting interleaved into the leapfrog algorithm gives the following iteration
\begin{equation} \label{eq:DR-iteration}
    \begin{split}
        \textbf{u}^{(j+1/2, j)} &= \mathcal{S}_{\text{ode}} \bigg(\frac{\Delta t}{2}, \textbf{u}'^{j}_0, \textbf{u}^{(j, j)} \bigg) \\
        \textbf{u}^{(j+1/2, j+1)} &= \mathcal{S}_{\text{pde}} \bigg(\Delta t, \textbf{u}'^{j}_0, \textbf{u}^{(j+1/2, j)} \bigg) \\
        \textbf{u}^{(j+1, j+1)} &= \mathcal{S}_{\text{ode}} \bigg(\frac{\Delta t}{2}, \textbf{u}'^{j}_0, \textbf{u}^{(j+1/2, j+1)} \bigg),
    \end{split}
\end{equation}
where $\mathcal{S}_{\text{ode}}$ is the solution to 
\begin{equation}
    \frac{d\textbf{u}}{dt} = \mathcal{A}_1(\textbf{u}) + \textbf{u}'_0,
\end{equation}
using  a time adaptive Adam-Bashforth method and $\mathcal{S}_{\text{pde}}$ represents the solution to
\begin{equation}
    \frac{d\textbf{u}}{dt} = \mathcal{A}_2(\textbf{u}) - \textbf{u}'_0,
\end{equation}
using a leapfrog staggered Crank-Nicolson time-centering algorithm. Analysis shows that SS-leapfrog+DR method is second-order accurate in time regardless of the choice of $\textbf{u}'_0$. This allows for $\textbf{u}'_0$ to be constructed to reduce the ODE/PDE splitting error. Our construction of $\textbf{u}'_0$ satisfies the following conditions:
\begin{itemize}
    \item $\textbf{u}'_0$ should vanish when $d\textbf{u}_{\text{pde}}/dt$ and $d\textbf{u}_{\text{ode}}/dt$ are additive, i.e. have same sign (this condition is expected to be rare and likely to only apply for fast dynamics where the operators are dominantly linear),
    \item $\textbf{u}'_0$ should be limited to the lesser of $d\textbf{u}_{\text{pde}}/dt$ and $-d\textbf{u}_{\text{ode}}/dt$ (when one operator dominates, the implicit fluid solves or multistep ODE method associated with the dominant operator should determine the bulk of the change in $\textbf{u}$). 
\end{itemize}
To satisfy these constraints we construct $\textbf{u}'_0$ with a \texttt{minmax} function defined as
\begin{equation} \label{u0prime}
  \textbf{u}'^{j}_0 = -\min{\left( \max{\left( - \frac{|\mathcal{A}_2 \left( \textbf{u}^{(j+1/2, j)} \right)|} {|\mathcal{A}_1 \left( \textbf{u}^{(j+1/2, j)} \right) |}, 0 \right)} , 1 \right)} \mathcal{A}_1 \left( \textbf{u}^{(j+1/2, j)} \right).
\end{equation}

In the implementation, the form of $\mathcal{A}_2 \left( \textbf{u} \right)$ in Eq.~\eqref{u0prime} is approximate for computational efficiency. With a $C_0$ finite-element discretization, calculation of the second-order derivatives associated with the PDE equations, $\mathcal{A}_2 \left( \textbf{u} \right)$, requires integration by parts and an additional mass-matrix solve. In order to avoid the cost associated with this solve, we use an approximation from the prior time-step solve

\begin{equation} \label{eq:u0p}
    \mathcal{A}_2 \left( \textbf{u}^{(j+1/2, j)} \right) \simeq \frac{\textbf{u}^{(j, j)} - \textbf{u}^{(j, j-1)}}{\Delta t} + \textbf{u}'^{j-1}_0
\end{equation}

\begin{figure}
    \centering
    \includesvg[scale=0.7]{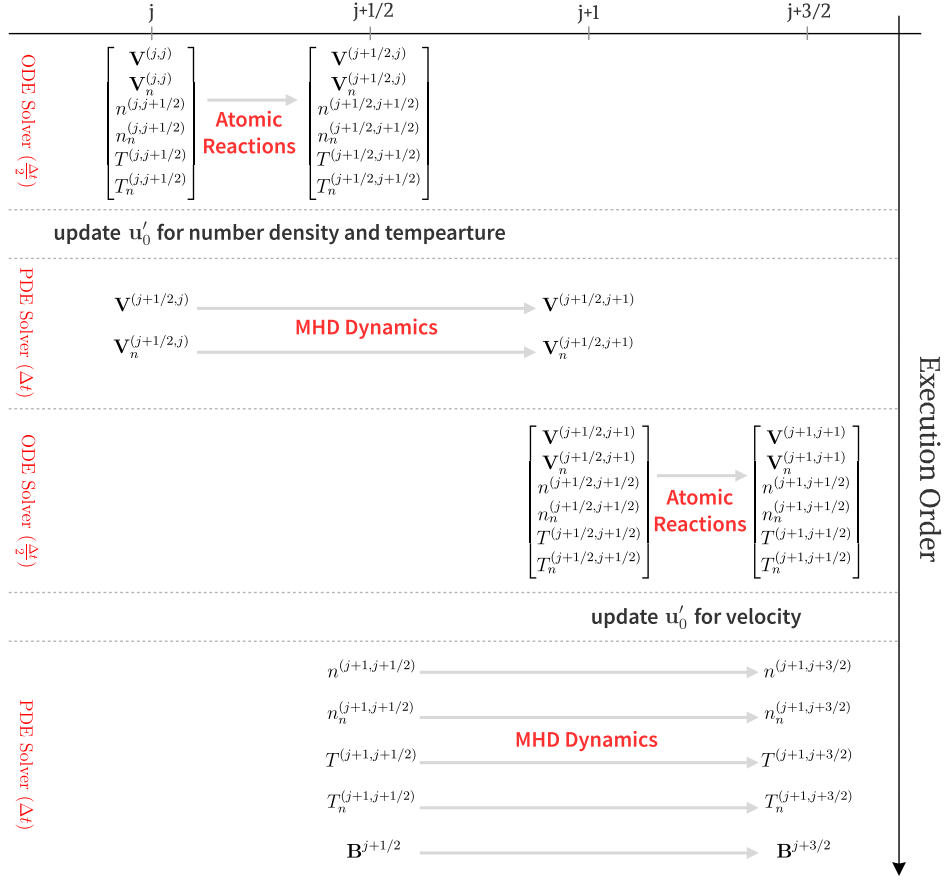}
    \caption{Schematic of SS-leapfrog+DR time advance. Strang-split is interleaved into the leapfrog integrating all the fields simultaneously, while the fields are staggered in the leapfrog algorithm and are solved independently.}
    \label{fig:SS-DU-scheme}
\end{figure}

Figure~\ref{fig:SS-DU-scheme} shows a schematic of the Strang-split time advance within the leapfrog with Douglas-Rachford inspired coupling. All the fields are solved simultaneously in the ODE steps while we still have the field staggering for the PDE solver. It is important to note that we update $\textbf{u}'_0$ for different fields at different spots in the iteration scheme to properly center the operator splitting with respect to the different fields in the leapfrog. The leapfrog advance for fields is staggered limiting the second superscript in Eq.~\eqref{eq:DR-iteration} to velocity advance only. For full description of the superscripts for fields consult Fig.~\ref{fig:SS-DU-scheme}.

\subsection{Discussion of the sources of the error} \label{subsec:error_source}
The prior discussion identifies multiple sources of error for the different time-discretization algorithms. The focus of this paper is on the discretization of atomic reactions and any other contribution of errors is neglected. These different sources are summarized as
\begin{itemize}
    \item \textbf{Nonlinear error} \\
    Explicit inclusion of nonlinear contributions of plasma-neutral interactions introduces a source of error that can be substantial. We eliminate this error by adding a Jacobian-free Newton-Krylov iteration to the Crank-Nicolson method or alternatively using an adaptive time-step size ODE solver in the Strang-split methods. 
    \item \textbf{Crank-Nicolson time-centering error} \\
    Time centering in the Crank-Nicolson method results in a second order in time-step size error. Using an adaptive Adam-Bashforth ODE solver in the Strang-split methods eliminates this error associated with the atomic interactions. 
    \item \textbf{\textit{T-in-n} error} \\
    As discussed in Sec.~\ref{subsec:CN-Linear} including the atomic interactions between plasma and neutrals breaks the data dependency of the MHD equations that is exploited by the leapfrog algorithm. Solving number density and temperature equations independently leads to an error from temperature-dependent cross-section terms in the density equation. This error is eliminated in the SS-leapfrog+DR algorithm by interleaving a Strang-split solve for the full system of equations into the leapfrog algorithm as described in Sec.~\ref{subsec:DR-multispecies}. 
    \item \textbf{Field staggering error} \\
    Atomic reaction terms in center of mass velocity and temperature equations, Eqs.~\eqref{eq:plasma_com_vel}-\eqref{eq:neutral_temp}, depend on relative velocity between the species and staggering the velocity in the leapfrog algorithm introduces a source of error in plasma and neutral temperature solutions. Similar to \textit{T-in-n} error, the SS-leapfrog+DR algorithm as discussed in Sec.~\ref{subsec:DR-multispecies} eliminates this error.
    \item \textbf{ODE/PDE splitting error} \\
    As discussed in Sec.~\ref{subsec:SS-multispecies} the Strang-split methods divide the equations into constituent ODE and PDE parts. When there are large cancellations between the ODE and PDE terms the Strang-split method is prone to significant error with oscillation in the solution between the different steps of each solve. These oscillations can lead to incorrect steady states. SS-leapfrog+DR with proper choice of $\textbf{u}'_0$ is designed to reduce these oscillations by limiting the changes during the ODE or PDE advance.
\end{itemize}

Separation of errors will be discussed in the context of 0D and 1D problems in Secs.~\ref{sec:test0-D} and~\ref{sec:test1-D}, respectively.

\section{Spatially Uniform Benchmarks} \label{sec:test0-D}
Having described several potential time-advance algorithms, we now compare their accuracy and efficiency by running a series of spatially uniform tests of the plasma-neutral interactions in regimes representing typical fusion devices. Not having any spatial gradients reduces the SS-leapfrog+DR method to an adaptive time-step-size ODE integrator for all equations in 0D and the error is simply set by the tolerance to the solver. Thus, we ignore this method in 0D cases. Spatially nonuniform cases will be considered in Sec.~\ref{sec:test1-D}. In the battery of 0D tests we have conditions that mimics the tokamak core, tokamak edge, nonlinear ionization and flow relaxation. Here we are purely testing the atomic physics and all spatial dependent terms are zero.

Plasma and neutral species interact with each other through three main processes: ionization, recombination, and charge exchange. For a spatially uniform system, number densities depend only on ionization and recombination processes and the equilibrium point is when ionization balances recombination. This equilibration point depends on electron temperature which depends on total energy of the system. We ignore the effective ionization energy, i.e. plasma is optically thick and radiation energy is trapped to keep the total energy conserved. Including the energy loss will cool the system continuously until it reaches the temperature of zero. In this section, we look at the accuracy and efficiency of the two algorithms in finding the equilibrium point for a plasma-neutral system. 

\begin{figure}
    \centering
    \includesvg[scale=0.75]{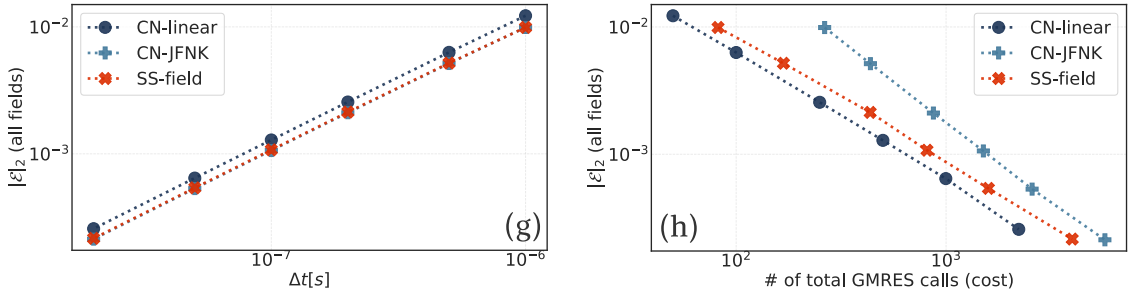}
    \caption{Accuracy and efficiency of methods for tokamak core (a, b), tokamak edge (c, d), nonlinear ionization (e, f), and flow relaxation (g, h) test cases. Total $l_2$ error is summation of $l_2$ errors for all changing fields normalized to a reference case at small time-step size. The number of GMRES iterations from the linear solver is a proxy for computational cost of each test case. Convergence plots (a, c, e, and g) represent the accuracy while cost plots (b, d, f, and h) show the efficiency of time discretization methods.}
    \label{fig:test-0-D}
\end{figure}

\subsection{Tokamak Core}
Consider a typical tokamak core condition with cold neutral fueling. At very high electron temperatures, the ionization rate is almost constant and recombination rarely happens. Thus, nonlinear terms in the number density equations are negligible compared to their linear counterparts in the source terms. High temperature plasma ionizes the neutral atoms and thus the plasma undergoes dilution cooling while the neutrals heat. Large changes in neutral temperature make the nonlinear source terms in the temperature equations dominant, and thus requires the full nonlinear solve. The initial states and changes in the plasma and neutral fields are given in Table~\ref{tab:tokamak-core}. Clear distinction between errors for different algorithms makes this case a good starting point. 

\begin{table}[ht]
    \centering
    \begin{tabular}{@{} r r r c r r @{}}\toprule
         & \multicolumn{2}{c}{Initial} & \phantom{abc} & \multicolumn{2}{c}{Change ($\Delta$)} \\ [0.5ex]
        \cmidrule{2-3} \cmidrule{5-6}
         & Number Density & Temperature && Number Density & Temperature \\ [0.5ex] 
         & $[\text{m}^{-3}]$ & [eV] && $[\text{m}^{-3}]$ & [eV] \\ [0.5ex]
        \midrule
        Plasma & $10^{20}$ & $1241.7$ && $4.15 \times 10^{15}$ & $0.1$ \\ [1ex]
        Neutral & $10^{16}$ & $1.2417$ && $-4.15 \times 10^{15}$ & $1072.3$ \\ [1ex]
        \bottomrule
    \end{tabular}
    \caption{Initial state and change of plasma and neutral species in a typical tokamak core regime. Hot plasma in the core ionizes the neutral atoms while they are heating up to reach thermal equilibrium with the plasma. The final states do not represent the equilibrium point but rather the state at $t = 2\times 10^{-7}$ s.}
    \label{tab:tokamak-core}
\end{table}

Figure~\ref{fig:test-0-D}(a) shows the normalized error with respect to chosen time-step for the CN-linear (blue circle), the CN-JFNK (orange plus), and the SS-field method (green cross). Errors are calculated compared to results for a fine-temporal-resolution computation with $\Delta t = 2 \times 10^{12}$ s. The overall normalized error is defined as 
\begin{equation} \label{eq:total_err}
    \| \mathcal{E} \|_2 = \frac{1}{N} \sum_j \sum_i \sqrt{ \left( \frac{u_i^j - u_{i,ref}^j}{u_{i,ref}^j} \right)^2 },
\end{equation}
where subscript $i$ represents the components of vector $\textbf{u}$, composed of evolving fields n, \textbf{V}, and T for both species plus $\textbf{B}$, superscript $j$ indicates the time slice, and $N$ is the total number of outputs. Results are written in periodic manner where the time-intervals are 10th of the entire duration of each run. 

Comparing errors for different discretization schemes disentangles three different sources of error as identified in Sec.~\ref{subsec:error_source}. For instance different errors in Fig.~\ref{fig:test-0-D}(a) are indicated as:
\begin{itemize}
    \item Nonlinear error represented by the difference between CN-linear and CN-JFNK,
    \item Crank-Nicolson time centering error shown as the difference between CN-JFNK and SS-field,
    \item \textit{T-in-n} error demonstrated by the difference between SS-field and the tolerance to the solver which is set to $10^{-12}$.
\end{itemize}
The different errors are not strictly additive and there are cases with fortuitous cancellations. 

Solving the linear system using generalized minimal residual method (GMRES) in the semi-implicit leapfrog method is a time consuming section and we use the total number of GMRES iterations as a proxy to measure the cost for each method. Figure~\ref{fig:test-0-D}(b) shows the normalized error versus total number of linear solve iterations for different time-steps for the CN-linear (blue circle), the CN-JFNK (orange plus), and the SS-field method (green cross). We are using this cost proxy instead of wall clock time because none of the codes are optimized and there might be some deviation from these results after optimizing the codes. 

\subsection{Tokamak Edge}
In the edge region of tokamaks, low plasma temperature in combination with higher neutral number density enhances the atomic physics. Thus, the numerical algorithm gets more challenged. The ionization rate changes drastically at electron temperatures of around $1-5$~eV and recombination has a comparable contribution in atomic processes. Heat exchange with cold neutrals cools down the plasma which causes the recombination rate to become more dependent on number density. Since the change in number densities are tiny, nonlinear source terms have a negligible effect in the number-density advance. However, nonlinearities in the temperature equations have a tangible contribution to the error due to large changes in the plasma temperature. The initial states and changes in different fields for both species are given in Table~\ref{tab:tokamak-edge}. Here, the final state of computation is not exactly the steady state because we want to focus on a dynamic part of the solution.

\begin{table}[ht]
    \centering
    \begin{tabular}{@{} r r r c r r @{}}\toprule
         & \multicolumn{2}{c}{Initial} & \phantom{abc} & \multicolumn{2}{c}{Change ($\Delta$)} \\ [0.5ex]
        \cmidrule{2-3} \cmidrule{5-6}
         & Number Density & Temperature && Number Density & Temperature \\ [0.5ex] 
         & $[\text{m}^{-3}]$ & $[\text{eV}]$ && $[\text{m}^{-3}]$ & $[\text{eV}]$ \\ [0.5ex]
        \midrule
        Plasma & $10^{17}$ & $2.4834$ && $1.64 \times 10^{14}$ & $-1.331$ \\ [1ex]
        Neutral & $10^{20}$ & $0.025$ && $-1.64 \times 10^{14}$ & $0.0027$ \\ [1ex]
        \bottomrule
    \end{tabular}
    \caption{Initial state and change for plasma and neutral species in a typical tokamak edge regime. The plasma loses energy by ionizing the cold neutrals in the process. The final states do not represent the equilibrium point but rather are the state at $t = 2\times 10^{-6}$~s.}
    \label{tab:tokamak-edge}
\end{table}

Figure~\ref{fig:test-0-D}(c) presents the summation of a normalized error for all changing fields versus time-step for different algorithms. The time-step size for reference computation for error calculation is $\Delta t = 2 \times 10^{-11}$~s. Since atomic cross sections are highly nonlinear dropping the nonlinear terms introduces error regardless of the size of the time-step. At large time-steps Crank-Nicolson time-centering error explains the difference between CN-JFNK and SS-field method, however, at smaller time-steps this error becomes negligible. Similar to the tokamak core case, \textit{T-in-n} error is presented by the difference between SS-field and solver error. Figure~\ref{fig:test-0-D}(d) shows the tipping point between CN-linear and CN-JFNK after which we get better accuracy relative to the computational cost by including the full nonlinearities in the solver.

\subsection{Nonlinear Ionization}
The initial state of the tokamak edge test case is close to equilibrium and number-density changes are small, thus the nonlinearities in the density equation are negligible despite the fact that the cross-section terms are highly nonlinear at low electron temperatures. In order to stress the algorithms further, a hypothetical regime is considered: nonlinear ionization where the initial state given in Table~\ref{tab:nonlinear_ion} is far from equilibrium and the change during the computations is considerable. The higher ionization rate results in increasing plasma number density and dilution cooling. Heat exchange between the species cools down the neutrals and keeps the total energy of the system constant (the ionization-energy loss is ignored here). The final state is not the asymptotic steady state but temperatures are largely equilibrated between species.  

\begin{table}[ht]
    \centering
    \begin{tabular}{@{} r r r c r r @{}}\toprule
         & \multicolumn{2}{c}{Initial} & \phantom{abc} & \multicolumn{2}{c}{Change ($\Delta$)} \\ [0.5ex]
        \cmidrule{2-3} \cmidrule{5-6}
         & Number Density & Temperature && Number Density & Temperature \\ [0.5ex] 
         & $[\text{m}^{-3}]$ & [eV] && $[\text{m}^{-3}]$ & [eV] \\ [0.5ex]
        \midrule
        Plasma & $10^{20}$ & $2.4834$ && $2.77 \times 10^{19}$ & $-0.2112$ \\ [1ex]
        Neutral & $10^{20}$ & $2.4834$ && $-2.77 \times 10^{19}$ & $-0.2111$ \\ [1ex]
        \bottomrule
    \end{tabular}
    \caption{Initial state and change for plasma and neutral species in a highly nonlinear regime. Ionization and recombination cross-sections change significantly with temperature, stressing \textit{T-in-n} error. Final states do not represent the equilibrium point but rather are the state at $t = 2\times 10^{-5}$~s.}
    \label{tab:nonlinear_ion}
\end{table}

Similar to the tokamak core and tokamak edge cases, Fig.\ref{fig:test-0-D}(e) shows the normalized error versus time-step for different algorithms. Time-step size for the reference computation is $\Delta t = 2 \times 10^{-10}$~s. Interestingly, the CN-linear method performs better than both the CN-JFNK and SS-field solvers. The different sources of error are not additive in this instance and there is partial cancellation of the errors. Diving more into the computation details, the high ionization rate results in large source terms in the density equations. Without nonlinear terms, the plasma number density solution overshoots, the higher plasma density enhances dilution cooling which in turn drops the ionization rate. Finally, the reduced ionization rate limits the source term and the overshoot in subsequent steps which partially cancels the initial error. The nonlinear algorithms eliminate the overshoot by including the nonlinear contributions to the source terms. However, the overall error is increased without partial cancellation of the \textit{T-in-n} error. The error cancellation in CN-linear is fortuitous for this specific case. 

Figure~\ref{fig:test-0-D}(f) shows the total number of GMRES iterations for each algorithm as a measure of computational cost. Error cancellation in the CN-linear method suggests that it is more efficient to ignore nonlinear terms in CN algorithms, however, this is an accidental coincidence. Although the SS-field method is the best of three, leapfrog staggering results in a considerable \textit{T-in-n} error which is eliminated by the Strang-split within the leapfrog.

\subsection{Flow Relaxation}
Up until now all the test cases are focused on number density and temperature equations. The flow relaxation test case examines the atomic reactions within the center-of-mass-velocity equation in conjunction with the atomic processes in the density and temperature. To focus on the velocity equation, plasma and neutral species start from a density equilibrium point where ionization and recombination rates balance and temperatures equilibrated. The plasma has a uniform velocity and neutrals are stationary. 

The high-velocity plasma exchanges momentum and energy with neutral particles through atomic reactions. In the process, the neutrals are ionized, heated, and dragged with plasma. The plasma and neutral thermal energies increase from frictional heating while plasma undergoes dilution cooling. The charge exchange time scales are much shorter than the ionization and recombination time scales and the number-density change for the species is negligible. Thus, the momentum and thermal equilibration dominates the dynamics reducing the dilution cooling. Table~\ref{tab:flow_relaxation} presents the initial state and changes in the fields for both species. The normalized error is calculated compared to a reference computation with $\Delta t = 1 \times 10^{-10}$~s.

\begin{table}[ht]
    \centering
    \begin{tabular}{@{} r r r r c r r r @{}}\toprule
         & \multicolumn{3}{c}{Initial} & \phantom{a} & \multicolumn{3}{c}{Change ($\Delta$)} \\ [0.5ex]
        \cmidrule{2-4} \cmidrule{6-8}
         & Number Density & Temperature & Velocity && Number Density & Temperature & Velocity \\ [0.5ex] 
         & $[\text{m}^{-3}]$ & [eV] & [m/s] && $[\text{m}^{-3}]$ & [eV] & [m/s] \\ [0.5ex]
        \midrule
        Plasma & $10^{19}$ & $1.2417$ & $10^4$ && $1 \times 10^{15}$ & $0.1141$ & $-5447$ \\ [1ex]
        Neutral & $1.492 \times 10^{19}$ & $1.2417$ & $0$ && $-1 \times 10^{15}$ & $0.1239$ & $3651$ \\ [1ex]
        \bottomrule
    \end{tabular}
    \caption{Initial state and change for plasma and neutral species in a flow relaxation test case. Plasma flow transfers momentum to neutral species through charge-exchange collisions as well as ionization until equilibrium is reached. The final states do not represent the equilibrium point but rather are the state at $t = 1\times 10^{-5}$~s.}
    \label{tab:flow_relaxation}
\end{table}

Figure~\ref{fig:test-0-D}(g) presents the error versus time-step size for all three methods. Among all the error sources, nonlinear contributions have a slight contribution to the total error and the field staggering is the dominant source of error. Since the number density and temperature fields are close to equilibrium through all the run time, \textit{T-in-n} error is negligible. Crank-Nicolson time-centering has no tangible effect in the total error. Similar to previous cases, Fig.~\ref{fig:test-0-D}(h) shows the total error as a function of computational cost.

\section{Spatially Non-uniform Benchmarks} \label{sec:test1-D}
In this section, we move beyond 0D cases and analyze the effects of spatially dependent MHD and fluid dynamics in conjunction with atomic reactions. CN-linear as a linear-implicit method is not robust for all 1D test cases and can give spurious results at large time-step sizes. Therefore, we drop the CN-linear method from analysis and focus on CN-JFNK as an implicitly-balanced method and compare it to the Strang-split methods. 

\subsection{Ballistic Expansion into Tokamak Pedestal}
\begin{figure}
    \centering
    \includesvg[scale=0.6]{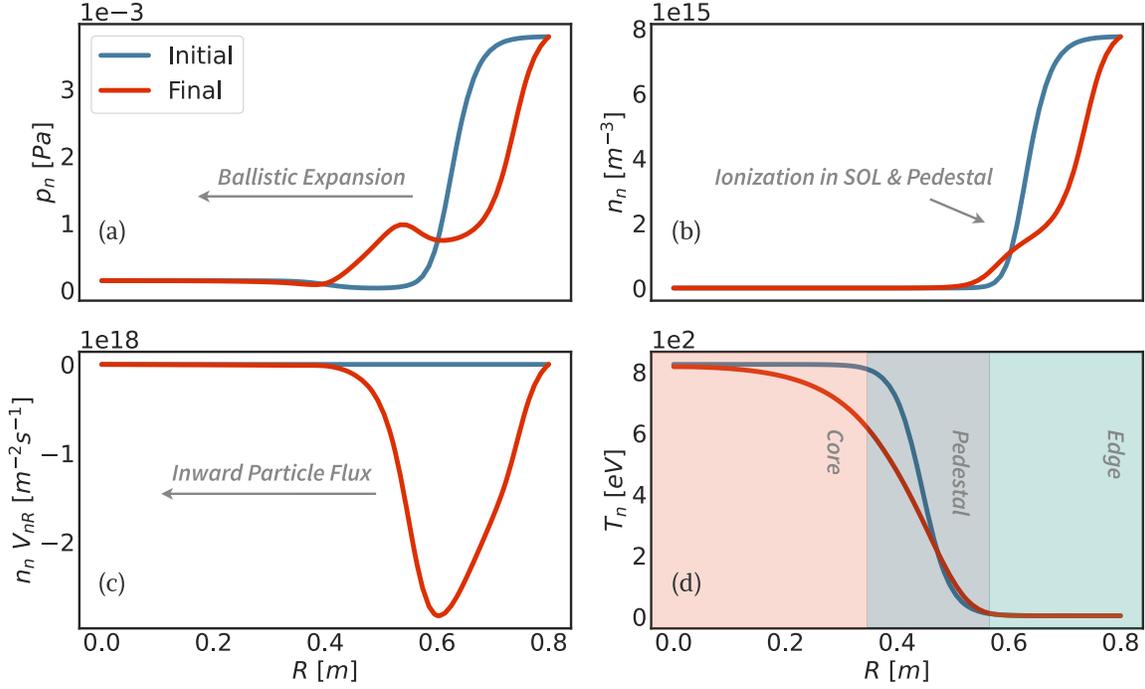}
    \caption{Initial and final ($t=0.5$ ms) profiles for (a) pressure, (b) number density, (c) number density flux, and (d) temperature of the neutral species in 1D tokamak pedestal case.}
    \label{fig:pedestal}
\end{figure}

Fusion-plasma edge fueling is one of the main applications of plasma-neutral MHD models. We investigate the spatially dependent neutral profiles in 1D with a generic tokamak pedestal test. Plasma profiles are kept stationary throughout the simulation and neutral initial state is set by a local equilibration of ionization and recombination as shown by red lines in Fig.~\ref{fig:pedestal}. To maintain a source of neutral particles during the steady state a thin diffusive layer and Dirichlet boundary conditions are applied to the number-density equation at the wall. Isotropic viscosity and heat conduction is assumed for neutrals with diffusivity of $0.1$~$\text{m}^2 \text{s}^{-1}$ and $0.01$~$\text{m}^2 \text{s}^{-1}$ at $10^{19}$~$\text{m}^{-3}$, respectively. The neutral pressure gradient causes ballistic expansion of particles into the core that is balanced by ionization. At the time-asymptotic steady state, shown with blue lines in Fig.~\ref{fig:pedestal}, the plasma is ionized in the edge pedestal before reaching the core. We analyze both the steady-state and dynamic solutions with the different time-advance algorithms. 

\subsubsection{Steady-State Solution}
\begin{figure}
    \centering
    \includesvg[scale=0.75]{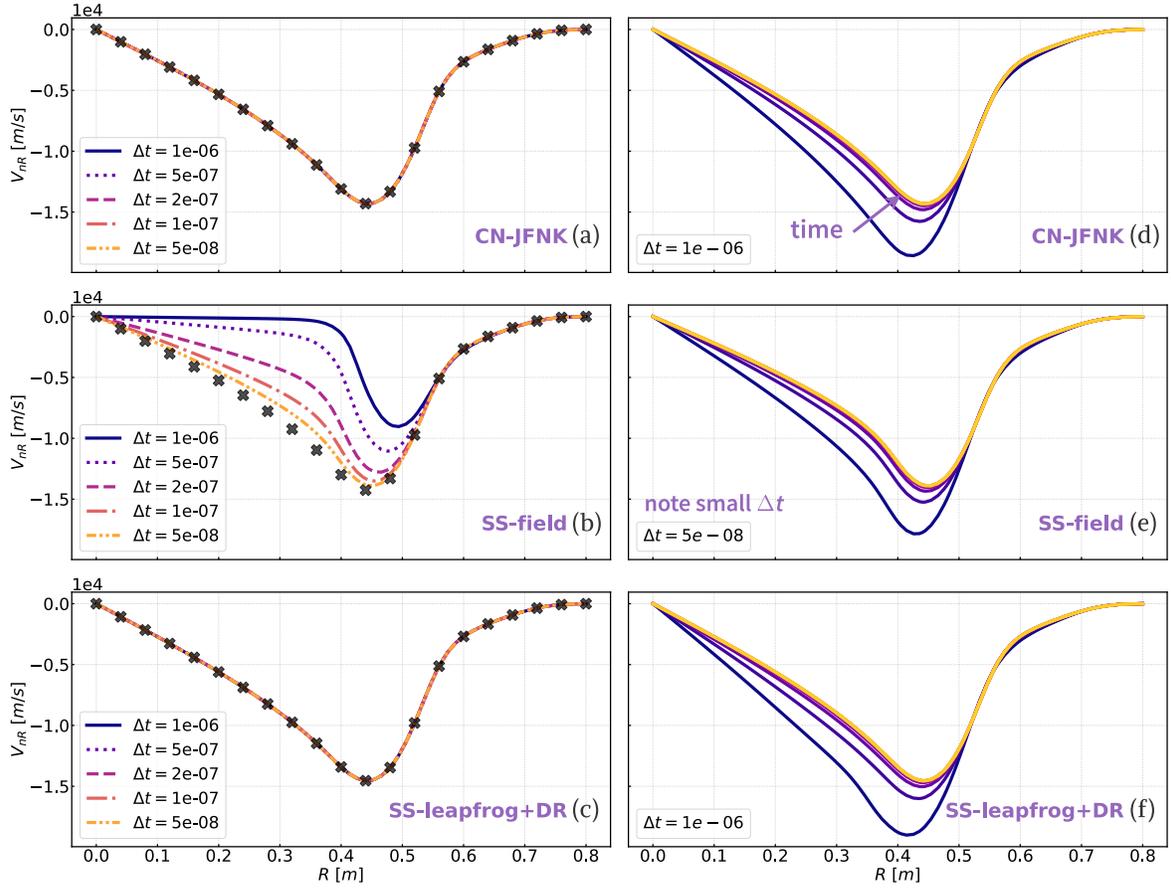}
    \caption{Steady state (a-c) and time dynamic (d-f) velocity profiles for neutral atoms in the tokamak pedestal case. ODE/PDE splitting error is significantly affecting the SS-field profiles (b), where using SS-leapfrog+DR (c) reduces the error to levels comparable to CN-JFNK (a). Similarly, SS-leapfrog+DR (f) follows CN-JFNK (d) in capturing the dynamic evolution of neutral velocity profile.}
    \label{fig:pedestal_results}
\end{figure}

Figures~\ref{fig:pedestal_results}(a-c) show the steady state profiles for three time-advance algorithms using different time-step sizes. Time-centering the atomic processes with Crank-Nicolson algorithm converges rapidly as seen in Fig.~\ref{fig:pedestal_results}(a). However, the ODE/PDE splitting error has a significant role in the SS-field solution and the numerical steady state is dependent on the time-step size~\cite{LeVeque2002}. SS-leapfrog+DR removes the ODE/PDE splitting error through the coupling parameter $\textbf{u}'_0$. For SS-leapfrog+DR, simultaneously solving the fields in the ODE system removes the \textit{T-in-n} error, however, with fixed plasma profiles the \textit{T-in-n} error has no effect.  

As discussed in Sec.~\ref{subsec:SS-multispecies} splitting the operators in the SS methods separates the PDE equations for the plasma and neutral species. This independent solution of fields makes the underlying linear system smaller in size and the associated GMRES solves computationally cheaper compared to the CN-JFNK method which couples the species within the linear solves. Unlike the 0D cases, 1D cases require periodic matrix refactorizations which makes using only the GMRES iteration count a poor proxy for cost. Relative to SS-field, SS-leapfrog+DR has slightly faster ODE solves. Comparing the wall clock time, the CN-JFNK method takes approximately three times more than the SS-field method with the same time-step size, while SS-field takes about 10 percent longer relative to SS-leapfrog+DR. These simulations are relatively low resolution (they use a finite-element grid with $40 \times 4$ bi-quintic elements), and we expect that the SS methods will better outperform the CN on larger system sizes.

\begin{figure}
    \centering
    \includesvg[scale=0.3]{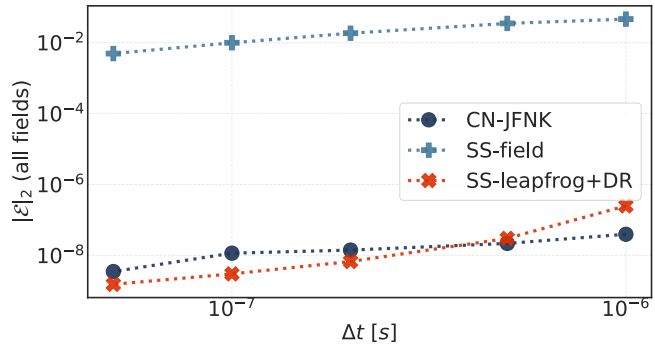}
    \caption{Accuracy and convergence of the different time-discretization methods for tokamak pedestal case. The y-axis represents the summation of $l_2$ errors for all neutral fields. The errors are calculated compared to a base solution with $\Delta t = 5 \times 10^{-9}$~s. The CN-JFNK and SS-leapfrog+DR deliver comparable accuracy.}
    \label{fig:pedestal_convergence}
\end{figure}

Figure~\ref{fig:pedestal_convergence} shows the convergence rate for all three methods at steady state. $l_2$-norm on the y axis is the summation of errors for neutral fields ($n_n, \textbf{V}_n$, and $T_n$) over the entire physical domain. SS-field method underestimates the neutral flux due to ODE/PDE splitting error, while the CN-JFNK and SS-leapfrog+DR have comparable accuracy.

\subsubsection{Dynamic Solution}
The dynamic solution often involves slow evolution resulting from the balance of large competing terms. These terms can be split between the ODE and PDE systems such as a balance between advection and ionization in the density equation and charge-exchange and the pressure gradient force. This case starts from a local atomic equilibrium and thus spans from a PDE dominated system initially to a balanced final state. There are more details in King et al.~\cite{King2021}.

\begin{figure}
    \centering
    \includesvg[scale=0.3]{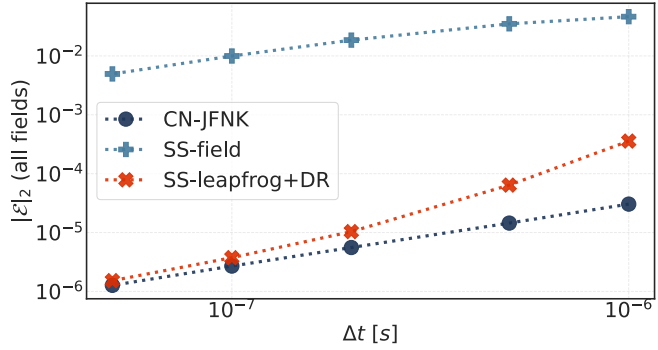}
    \caption{Time averaged error over all neutral fields in the tokamak pedestal case. During the whole simulation time ($0.5$~ms), profiles are compared to a temporally fine-resolved base-solution on a periodic basis (every 10 percent of the duration). The averaged error for SS-leapfrog+DR is comparable to the CN-JFNK, while SS-field retains noticeable error.}
    \label{fig:pedestal_error_avg}
\end{figure}

Figures~\ref{fig:pedestal_results}(d-f) show the velocity profile of the tokamak pedestal test case for 10 intermediate states (starting with red) until we reach the asymptotic steady state (blue line) for all three methods. At the beginning the neutral flux to the core is very large which slows down as it reaches the high-temperature, dense plasma in the core. High charge-exchange rate as well as ionization acts as a balancing force to pressure gradient which drives the flux. As it reaches steady state a flux of neutral particles move from wall to the core as is shown in Fig.~\ref{fig:pedestal}(c). The time-dynamic evolution of all three algorithms, Figs.~\ref{fig:pedestal_results}(d-f), is qualitatively similar (but not exactly and note the small time-step size for Fig.~\ref{fig:pedestal_results}(e)). Both SS algorithms underpredict the core penetration velocity, but this is particularly prevalent for the SS-field algorithm. As shown in Fig.~\ref{fig:pedestal_error_avg}, by reducing the time-step size the time averaged error for SS-leapfrog+DR becomes comparable to the CN-JFNK. 

\subsection{Planar Electromagnetic Plasma Accelerator}
\begin{figure}
    \centering
    \includesvg[scale=0.65]{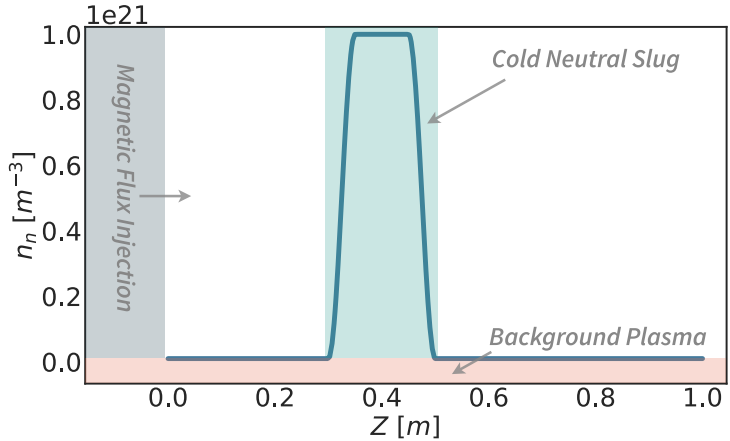}
    \caption{Initial neutral number density profile for snowplow case sitting in a uniform background plasma. A linearly rising magnetic flux is injected from left boundary pushing plasma towards the cold neutral slug.}
    \label{fig:snowplow}
\end{figure}

Next, we present a time-dynamic test case where full interaction between plasma and neutral species is studied. A 1D snowplow includes dominantly hyperbolic and atomic reactions when a slug of cold neutrals is hit by an accelerating plasma. A linearly rising magnetic flux is injected from the left side into an infinitely long domain with uniform background plasma as shown in Fig.~\ref{fig:snowplow}. This setup is a simplified representation of the acceleration region in a sheared-flow-stabilized Z-pinch. Calculations are done using a finite-element grid with $64 \times 4$ bi-quintic elements on a 2-D domain. Considering a periodic boundary condition in one of the directions reduces the case to a 1D plasma acceleration. Background plasma has a uniform number density of $10^{19}$~$\text{m}^{-3}$. Similarly, neutral number density is $10^{19}$~$\text{m}^{-3}$ everywhere except in the slug where it is $10^{21}$~$\text{m}^{-3}$. Perpendicular component of magnetic field at the boundary rises to $0.02$~T in $10$ $\mu$s. Plasma resistivity is assumed to be constant with a value of $100 \mu_0$~$\Omega \cdot$m. We use isotropic viscosity and thermal conduction for both plasma and neutral species. Plasma viscosity (diffusivity of $400$~$\text{m}^2 \text{s}^{-1}$ at $10^{19}$~$\text{m}^{-3}$) and thermal conductivity (diffusivity of $500$~$\text{m}^2 \text{s}^{-1}$ at $10^{19}$~$\text{m}^{-3}$) are calculated following Braginskii. Finally, neutral viscosity (diffusivity of $400$~$\text{m}^2 \text{s}^{-1}$ at $10^{19}$~$\text{m}^{-3}$) and thermal conductivity (diffusivity of $500$~$\text{m}^2 \text{s}^{-1}$ at $10^{19}$~$\text{m}^{-3}$) are based on the Sutherland formula.

\begin{figure}
    \centering
    \includesvg[scale=1.0]{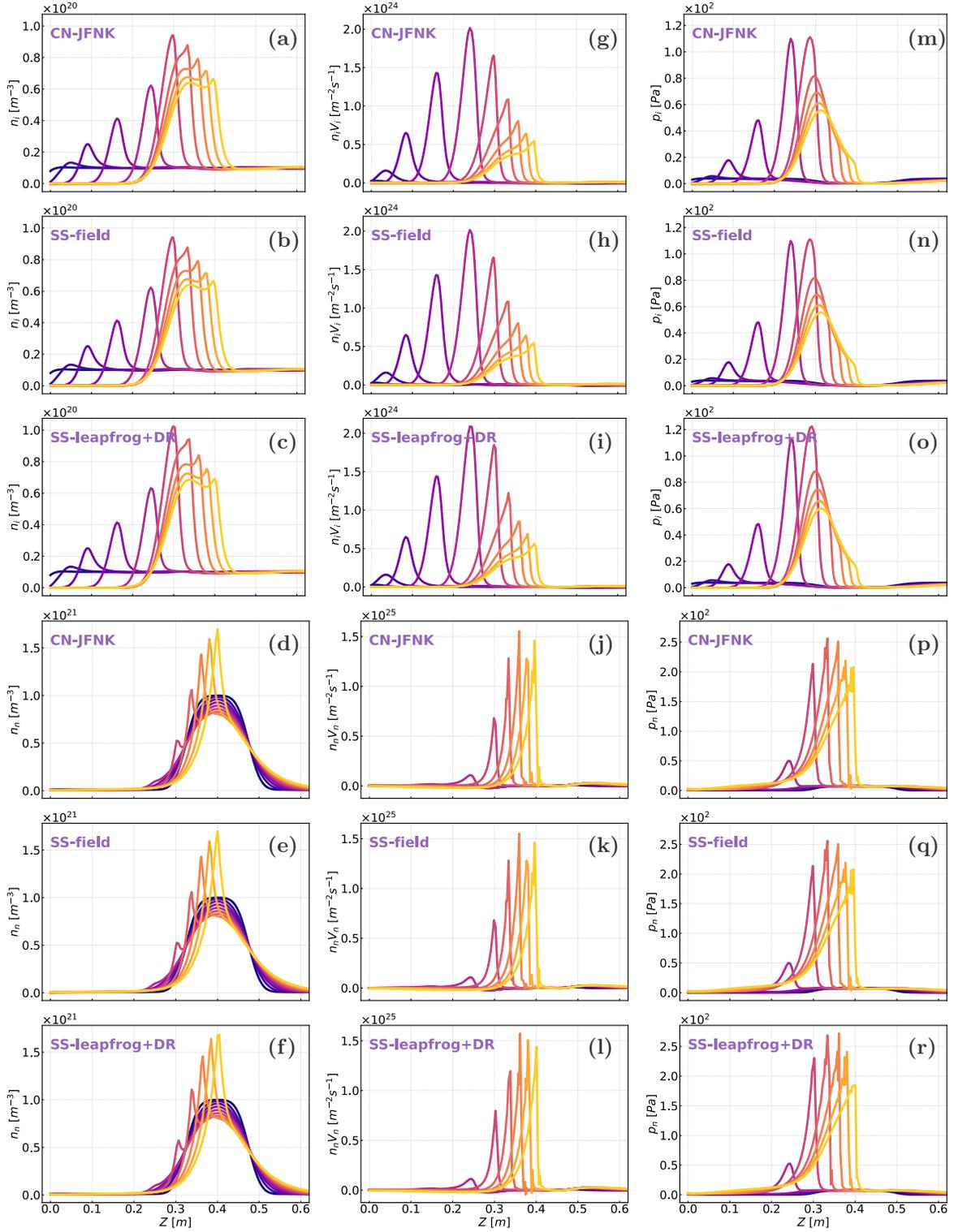}
    \caption{Dynamic evolution of the plasma and neutral species in a planar electromagnetic plasma accelerator with $\Delta t = 10^{-8}$~s. Top three rows show the plasma number density, plasma momentum, and plasma pressure. Bottom three rows show the same variables for neutral species. SS-field delivers comparable accuracy when ODE and PDE operators have the same sign.}
    \label{fig:sp_dynamic}
\end{figure}

\begin{figure}
    \centering
    \includesvg[scale=0.3]{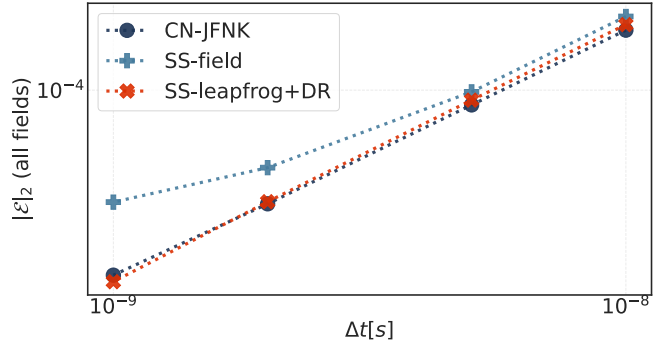}
    \caption{Accuracy and convergence of the different time-sidcretization methods for generic snowplow case. $l_2$-error on the y-axis is the summation of errors for all fields compared to a base solution with $\Delta t = 10^{-10}$~s.}
    \label{fig:sp_convergence}
\end{figure}

Figure~\ref{fig:sp_dynamic} shows the dynamic evolution of the number density, momentum, and pressure for both species (starting with dark color) up to 20 $\mu s$ with $\Delta t = 10^{-8}$~s. Rising magnetic flux accumulates magnetic pressure until it surpasses the thermal pressure of the plasma and creates a magnetic piston pushing the plasma to the right. According to snowplow theory, the magnetic piston is led by a shock wave where heated plasma starts to ionize the neutral slug upon reaching it which increases the plasma number density. Figures~\ref{fig:sp_dynamic}~(m-o) show a high pressure pick building up in plasma which starts to widen when it hits the neutral slug. Widening of the plasma number-density profile in Figs.~\ref{fig:sp_dynamic}~(a-c) represents the neutral ionization. In the process, plasma loses momentum due to ionization and charge exchange collisions which drags the neutrals with plasma which can be seen in Figs.~\ref{fig:sp_dynamic}~(g-l). Magnetic flux at the left boundary is implemented as a time-dynamic Dirichlet condition on the perpendicular component of the magnetic field. A zero Neumann boundary condition is used for number-density and velocity equations, where it acts as an open-flow boundary on velocity equation. All three methods deliver comparable accuracy in solving dynamic problems as can be seen in Fig.~\ref{fig:sp_convergence}. Unlike the tokamak pedestal test case, SS-field method gives fairly accurate results here which indicates that the ODE/PDE splitting error is not dominant. 

Figure~\ref{fig:sp_convergence} presents the time-averaged $l_2$ error of all fields over the entire physical domain. The error is calculated compared to a base solution with $\Delta t = 10^{-10}$~s. CN-JFNK and SS-leapfrog+DR methods deliver comparable accuracy and convergence while the SS-field method converges more slowly and is less accurate at small time-step sizes. We conclude that for this case there is a common source of error for all three methods at large time-steps, likely from the PDE Crank-Nicolson and leapfrog discretization itself. At small time-step size, the PDE/ODE splitting error becomes dominant in SS-field method and adversely affects convergence.

\section{Conclusion} \label{sec:conclusion}
The goal of this work is to evaluate the best time-discretization algorithm compatible with a semi-implicit leapfrog for inclusion of atomic interactions in a plasma-neutral MHD model. Two candidate approaches are considered: Crank-Nicolson with JFNK (CN-JFNK) and Strang-split which is further subdivided into split-by-field (SS-field) and leapfrog-interleaved with Douglas-Rachford-inspired coupling (SS-leapfrog+DR). The Strang-split algorithms have many advantages. Testing on 0D cases, Sec.~\ref{sec:test0-D}, shows the ODE solver accurately eliminates both the need for nonlinear iteration when solving the highly nonlinear atomic-interactions equations and the CN time-centering error associated with the atomic terms (the fluid CN centering error is still present in the SS algorithm, but this error is not present in 0D). Furthermore, the SS-leapfrog+DR method eliminates errors associated with the combination of the leapfrog and atomic interactions, identified as \textit{T-in-n} and field-staggering error is Sec.~\ref{subsec:error_source}. In fact, this algorithm reduces the error to that of the multi-step ODE solver tolerance for 0D cases and thus it is not considered in our 0D tests.

Consideration of tests with spatially dependent fields in Sec.~\ref{sec:test1-D} introduces ODE/PDE splitting error when using the Strang-split methods. We consider two examples: fusion-plasma edge fueling and a co-axial plasma-injector snowplow which are representative of slow and fast applications, respectively. Through the use of a Douglas-Rachford-inspired coupling, the ODE/PDE splitting error is minimized in SS-leapfrog+DR to be comparable to CN-JFNK. In fact, analysis of the accuracy shows that the error associated with SS-leapfrog+DR can be lower than CN-JFNK at the same time-step size as interleaving the ODE solves with the leapfrog eliminates the \textit{T-in-n}, field-staggering and CN error associated with the atomic interaction terms. 

In addition, the Strang-split methods have a robustness advantage relative to CN-JFNK. Running permutations of the snowplow case requires modification of the nonlinear solver tolerances and some cases fail to converge to a satisfactory solution potentially because our JFNK is approximate.

Finally, the Strang-split methods have a computational performance advantage relative to CN. Each species is considered independently during global-matrix solves. This significantly reduces the algebraic system size and fill-in and permits exploitation of parallelization over species (which is not presently explored). The test cases presented are relatively small but the performance advantage is evident. Given the scaling of global-matrix solves, we expect the performance advantage of the Strang-split methods to become larger with high-resolution cases. The advantages only grow when considering multiple species, such as tritium, carbon, and tungsten, where parallelization over species would lead to a huge performance gain. Future work will involve application of the SS-leapfrog+DR to this system of equations.

\section*{Acknowledgments}

We thank Eric Howell for stimulating discussions on topics related to this work.
We thank Eric Howell and Scott Kruger for proof-reading the draft manuscript. 
Sina Taheri executed the bulk of this work including drafting the manuscript,
design and implementation of the Crank-Nicolson algorithms and designed and executing the
test cases. Jacob King designed and implemented the Strang-Split algorithms.

This material is based on work supported by the U.S.  Department of Energy
Office of Science under contract numbers DE-SC0016256 (Taheri and Shumlak) 
and DE-SC0018311 (King and Taheri).

\begin{appendices}
\section{Discretized system of equation with Crank-Nicholson time centering}
\label{sec:appendixA}
\begin{equation}\label{eq:app_vcom}
\begin{split}
    m n^{j + \frac{1}{2}} &\left( \frac{\Delta \textbf{V}}{\Delta t} + \theta_v \textbf{V}^j \cdot \nabla \Delta \textbf{V} + \theta_v \Delta \textbf{V} \cdot \nabla \textbf{V}^j + \theta_v^2 \Delta \textbf{V} \cdot \nabla \Delta \textbf{V} \right) - \Delta t \textbf{L}^{j+\frac{1}{2}}(\Delta \textbf{V}) + \frac{1}{2} \nabla \cdot \underbar{$\Pi$} (\Delta \textbf{V}) \\	
    &- \textbf{R}_{in}^{cx, j+\frac{1}{2}}(\Delta \textbf{V}, \textbf{V}_{n}^{j}) + \textbf{R}_{ni}^{cx, j+\frac{1}{2}}(\Delta \textbf{V}, \textbf{V}_{n}^{j}) \\
    &+ \theta_v m \Gamma^{ion} \Delta \textbf{V} + \theta_v m \Gamma^{cx}(\textbf{V}^{j}, \textbf{V}_{n}^{j}) \Delta \textbf{V} - m (\textbf{V}_{n}^{j} - \textbf{V}^{j}) \Gamma^{cx, j+\frac{1}{2}}(\Delta \textbf{V}, \textbf{V}_{n}^{j}) \\
	= &- m n^{j+\frac{1}{2}} \textbf{V}^{j} \cdot \nabla \textbf{V}^{j} +\textbf{J}^{j+\frac{1}{2}} \times \textbf{B}^{j+\frac{1}{2}} - \nabla p^{j+\frac{1}{2}} - \nabla \cdot \underbar{$\Pi$} (\textbf{V}^j) \\
	&+ \textbf{R}_{in}^{cx}(\textbf{V}^{j}, \textbf{V}_n^{j}) -  \textbf{R}_{ni}^{cx}(\textbf{V}^{j}, \textbf{V}_n^{j}) + m \Gamma^{ion} (\textbf{V}_n^{j} - \textbf{V}^{j}) + m \Gamma^{cx}(\textbf{V}^{j}, \textbf{V}_{n}^{j}) (\textbf{V}_n^{j} - \textbf{V}^{j})	
\end{split}
\end{equation}

\begin{equation}\label{eq:app_vneut}
\begin{split}
	m n_n^{j + \frac{1}{2}} &\left( \frac{\Delta \textbf{V}_n}{\Delta t} + \theta_v \textbf{V}_n^j \cdot \nabla \Delta \textbf{V}_n + \theta_v \Delta \textbf{V}_n \cdot \nabla \textbf{V}_n^j + \theta_v^2 \Delta \textbf{V}_n \cdot \nabla \Delta \textbf{V}_n \right) - \Delta t \textbf{L}_n^{j+\frac{1}{2}}(\Delta \textbf{V}_n) + \frac{1}{2} \nabla \cdot \underbar{$\Pi$}_n (\Delta \textbf{V}_n) \\
	&- \textbf{R}_{ni}^{cx, j+\frac{1}{2}}(\textbf{V}^{j}, \Delta \textbf{V}_n) + \textbf{R}_{ni}^{cx, j+\frac{1}{2}}(\textbf{V}^{j}, \Delta \textbf{V}_n) \\
	&+ \theta_v m \Gamma^{rec} \Delta \textbf{V}_{n} + \theta_v m \Gamma^{cx} (\textbf{V}^{j}, \textbf{V}_{n}^{j}) \Delta \textbf{V}_n - m (\textbf{V}^{j} - \textbf{V}_{n}^{j}) \Gamma^{cx, j+\frac{1}{2}} (\textbf{V}^{j}, \Delta \textbf{V}_n)\\
	= &- m n_n^{j+\frac{1}{2}} \textbf{V}_n^{j} \cdot \nabla \textbf{V}_n^{j} - \nabla p_n^{j+\frac{1}{2}} - \nabla \cdot \underbar{$\Pi$}_n(\textbf{V}_n^j) \\
	&+ \textbf{R}_{ni}^{cx}(\textbf{V}^{j}, \textbf{V}_n^{j}) - \textbf{R}_{in}^{cx}(\textbf{V}^{j}, \textbf{V}_n^{j}) + m \Gamma^{rec} (\textbf{V}^{j} - \textbf{V}_n^{j}) + m \Gamma^{cx} (\textbf{V}^{j}, \textbf{V}_{n}^{j}) (\textbf{V}^{j} - \textbf{V}_n^{j}) 
\end{split}
\end{equation}

\begin{equation}\label{eq:app_pdens}
\begin{split}
	\frac{\Delta n}{\Delta t} &+ \theta \nabla \cdot (\textbf{V}^{j+1} \Delta n) - \Gamma^{ion, j+1} (\Delta n, \Delta n_n) + \Gamma^{rec, j+1} (\Delta n) \\
	&= - \nabla \cdot (\textbf{V}^{j+1} n^{j+\frac{1}{2}}) + \Gamma^{ion}(n^{j+\frac{1}{2}}, n_n^{j+\frac{1}{2}}) - \Gamma^{rec}(n^{j+\frac{1}{2}})
\end{split}
\end{equation}

\begin{equation}\label{eq:app_ndens}
\begin{split}
	\frac{\Delta n_n}{\Delta t} &+ \theta \nabla \cdot (\textbf{V}_n^{j+1} \Delta n_n) - \Gamma^{rec, j+1} (\Delta n) + \Gamma^{ion, j+1} (\Delta n, \Delta n_n) \\
	&= - \nabla \cdot (\textbf{V}_n^{j+1} n_n^{j+\frac{1}{2}}) +\Gamma^{rec}(n^{j+\frac{1}{2}}) - \Gamma^{ion}(n^{j+\frac{1}{2}}, n_n^{j+\frac{1}{2}})
\end{split}
\end{equation}

\begin{equation}\label{eq:app_ptemp}
\begin{split}
	\frac{\bar{n}}{\Gamma - 1} & \left( \frac{\Delta T}{\Delta t} + \theta \textbf{V}^{j+1} \cdot \nabla \Delta T \right) + \theta \bar{n} \Delta T \nabla \cdot \textbf{V}^{j+1} + \theta_{visc} \nabla \cdot \textbf{q}(\Delta T) \\
	&+ \frac{1}{\Gamma - 1} \Delta T \left[ \Gamma^{ion}(T^{j+\frac{1}{2}}) - \Gamma^{rec}(T^{j+\frac{1}{2}}) \right] + \frac{1}{\Gamma - 1} T_{n}^{j+\frac{1}{2}} \left[ \Gamma^{ion, j+1}(\Delta T) - \Gamma^{rec, j+1}(\Delta T) \right] \\
	&- Q^{ion, j+1}(\Delta T, \Delta T_n) + Q^{rec, j+1}(\Delta T, \Delta T_n) - Q_{in}^{cx, j+1}(\Delta T, \Delta T_n) + Q_{ni}^{cx, j+1}(\Delta T, \Delta T_n) \\
	&- \frac{m}{2} \left[\Gamma^{ion}(\Delta T) + \Gamma^{cx, j+1}(\Delta T, \Delta T_n) \right] (\textbf{V}^{j+1} - \textbf{V}_{n}^{j+1})^2 + \textbf{R}_{in}^{cx, j+1}(\Delta T, \Delta T_n) \cdot (\textbf{V}^{j+1} - \textbf{V}_{n}^{j+1}) \\
	=& - \frac{\bar{n}}{\Gamma - 1} \textbf{V}^{j+1} \cdot \nabla T^{j+\frac{1}{2}} - \bar{n} T^{j+\frac{1}{2}} \nabla \cdot \textbf{V}^{j+1} - \nabla \cdot \textbf{q}(T^{j+\frac{1}{2}}) + Q^{j+\frac{1}{2}} \\
	&- \frac{1}{\Gamma - 1} T_{n}^{j+\frac{1}{2}} \left[ \Gamma^{ion}(T^{j+\frac{1}{2}}) - \Gamma^{rec}(T^{j+\frac{1}{2}}) \right] \\
	&+ Q^{ion}(T^{j+\frac{1}{2}}, T_{n}^{j+\frac{1}{2}}) - Q^{rec}(T^{j+\frac{1}{2}}, T_{n}^{j+\frac{1}{2}}) + Q_{in}^{cx}(T^{j+\frac{1}{2}}, T_{n}^{j+\frac{1}{2}}) - Q_{ni}^{cx}(T^{j+\frac{1}{2}}, T_{n}^{j+\frac{1}{2}}) \\
	&+ \frac{m}{2} \left[\Gamma^{ion}(T^{j+\frac{1}{2}}) + \Gamma^{cx}(T^{j+\frac{1}{2}}, T_{n}^{j+\frac{1}{2}}) \right] (\textbf{V} - \textbf{V}_{n})^2 - \textbf{R}_{in}^{cx}(T^{j+\frac{1}{2}}, T_{n}^{j+\frac{1}{2}}) \cdot (\textbf{V} - \textbf{V}_{n})
\end{split}
\end{equation}

\begin{equation}\label{eq:app_ntemp}
\begin{split}
	\frac{\bar{n}_n}{\Gamma - 1} & \left( \frac{\Delta T_n}{\Delta t} + \theta \textbf{V}_n^{j+1} \cdot \nabla \Delta T_n \right) + \theta \bar{n}_n \Delta T_n \nabla \cdot \textbf{V}_n^{j+1} + \theta_{visc} \nabla \cdot \textbf{q}_n(\Delta T_n) \\
	&+ \frac{1}{\Gamma - 1} \Delta T_{n} \left[ \Gamma^{rec}(T^{j+\frac{1}{2}}) - \Gamma^{ion}(T^{j+\frac{1}{2}}) \right] + \frac{1}{\Gamma - 1} T_{n}^{j+\frac{1}{2}} \left[ \Gamma^{rec, j+1}(\Delta T) - \Gamma^{ion, j+1}(\Delta T) \right] \\
	&- Q^{rec, j+1}(\Delta T, \Delta T_n) + Q^{ion, j+1}(\Delta T, \Delta T_n) - Q_{ni}^{cx, j+1}(\Delta T, \Delta T_n) + Q_{in}^{cx, j+1}(\Delta T, \Delta T_n) \\
	&- \frac{m}{2} \left[\Gamma^{rec}(\Delta T) + \Gamma^{cx, j+1}(\Delta T, \Delta T_n) \right] (\textbf{V}^{j+1} - \textbf{V}_{n}^{j+1})^2 - \textbf{R}_{ni}^{cx, j+1}(\Delta T, \Delta T_n) \cdot (\textbf{V}^{j+1} - \textbf{V}_{n}^{j+1}) \\
	=& - \frac{\bar{n}_n}{\Gamma - 1} \textbf{V}_n^{j+1} \cdot \nabla T_n^{j+\frac{1}{2}} - \bar{n}_n T_n^{j+\frac{1}{2}} \nabla \cdot \textbf{V}_n^{j+1} - \nabla \cdot \textbf{q}_n (T_n^{j+\frac{1}{2}}) + Q_n^{j+\frac{1}{2}} \\
	&- \frac{1}{\Gamma - 1} T_{n}^{j+\frac{1}{2}} \left[ \Gamma^{rec}(T^{j+\frac{1}{2}}) - \Gamma^{ion}(T^{j+\frac{1}{2}}) \right] \\
	&+ Q^{rec}(T^{j+\frac{1}{2}}, T_{n}^{j+\frac{1}{2}}) - Q^{ion}(T^{j+\frac{1}{2}}, T_{n}^{j+\frac{1}{2}}) + Q_{ni}^{cx}(T^{j+\frac{1}{2}}, T_{n}^{j+\frac{1}{2}}) - Q_{in}^{cx}(T^{j+\frac{1}{2}}, T_{n}^{j+\frac{1}{2}}) \\
	&+ \frac{m}{2} \left[\Gamma^{rec}(T^{j+\frac{1}{2}}) + \Gamma^{cx}(T^{j+\frac{1}{2}}, T_{n}^{j+\frac{1}{2}}) \right] (\textbf{V} - \textbf{V}_{n})^2 + \textbf{R}_{ni}^{cx}(T^{j+\frac{1}{2}}, T_{n}^{j+\frac{1}{2}}) \cdot (\textbf{V} - \textbf{V}_{n})
\end{split}
\end{equation}

\begin{equation}\label{eq:app_indc}
\begin{split}
	\frac{\Delta \textbf{B}}{\Delta t} - \theta \nabla \times (\textbf{V}^{j+1} \times \Delta \textbf{B}) + \theta_{visc} \nabla \times (\eta \Delta \textbf{J}) = \nabla \times (\textbf{V}^{j+1} \times \textbf{B}^{j+\frac{1}{2}}) - \nabla \times (\eta \textbf{J}^{j+\frac{1}{2}})
\end{split}
\end{equation}
where the semi-implicit operators for plasma and neutral velocity equation are given as
\begin{equation}\label{eq:app_psemi_imp}
\begin{split}
    \textbf{L}(\Delta \textbf{V}) &= C_0 \left\{ \frac{1}{\mu_0} [\nabla \times \nabla \times (\Delta \textbf{V} \times \textbf{B})] \times \textbf{B} + \textbf{J} \times \nabla \times (\Delta \textbf{V} \times \textbf{B}) + \nabla \left( \Delta \textbf{V} \cdot \nabla p + \frac{5}{3} p \nabla \cdot \Delta \textbf{V} \right) \right\} \\ &+ C_1 p_{nl} \nabla^2 \Delta \textbf{V}
\end{split}
\end{equation}

\begin{equation}\label{eq:app_nsemi_imp}
    \textbf{L}_n(\Delta \textbf{V}_n) = C_0 \nabla \left( \Delta \textbf{V}_n \cdot \nabla p_n + \frac{5}{3} p_n \nabla \cdot \Delta \textbf{V}_n \right) + C_1 p_{n,nl} \nabla^2 \Delta \textbf{V}_n
\end{equation}
where $C_0$ is a coefficient for the ideal-MHD force operator, and $C_1$ is a coefficient for the Laplacian part with $p_{nl}$ and $p_{n,nl}$ the 'nonlinear pressure', which is typically orders of magnitude smaller than the total pressure.
\end{appendices}

\bibliographystyle{unsrt}
\bibliography{bibliography.bib}

\end{document}